\documentclass[aps, prb, letterpaper, 10pt, twocolumn]{revtex4-1}
\usepackage[colorlinks=true,allcolors=red!80!blue]{hyperref}
\usepackage{amssymb}
\usepackage{mathptmx}
\usepackage{amsmath}
\usepackage{graphicx}
\usepackage{subfigure}
\usepackage{amsfonts}
\usepackage{xcolor}
\usepackage{epsfig}
\usepackage{color}
\usepackage{bbm}
\usepackage{bm}
\usepackage{tabularx}
\usepackage{multirow}
\usepackage{mathtools}
\usepackage{xfrac}
\DeclareMathAlphabet{\mathcal}{OMS}{cmsy}{m}{n}

\newcommand{\ket}[1]{\vert{#1}\rangle} 
\newcommand{\bra}[1]{\langle{#1}\vert}
\newcommand{\proj}[1]{\ket{#1}\!\bra{#1}}

\newcommand{\abs}[1]{\left|#1\right|}

\newcommand{\tr}[2]{\mathrm{Tr}_{#1}\!\left\{#2\right\}}
\newcommand{\hatd}[1]{\hat{#1}^{\dagger}}

\newcommand{\half}{\frac{1}{2}}

\newcommand{\gele}{\textsl{g}_{\rm e}}

\newcommand{\rhotd}{\rho_{\rm 2D}}

\begin{document}
\title{Dynamical quantum phase transition in a system of non-interacting bosons}

\author{Mehdi Abdi}
\email{mehabdi@gmail.com}
\affiliation{Department of Physics, Isfahan University of Technology, Isfahan 84156-83111, Iran}

\begin{abstract}
We study a Bose-Einstein condensate at the low energy limit and show that their collective dynamics exhibit interesting quantum dynamical behavior.
The system undergoes a dynamical quantum phase transition after a sudden quench into a properly distributed static force, provided its dispersion relation is linear.
We corroborate the occurrence of the dynamical phase transition by calculating Fisher zeros of the Loschmidt amplitude and showing that they cross the real time axis in thermodynamic limit.
A connection is established between the order of nonanalycity in the return rate function and the spectral density of the force.
Furthermore, it is shown that a logarithmic or power law scaling holds at the critical times depending on the displacement spectrum.
The scaling behaviors are studied for three different cases.
Eventually, a scheme for the quantum simulation of such dynamical phase transition and its verification is proposed.
We show that the behavior remains observable even by taking into account the environmental effects.
\end{abstract}

\maketitle

%
%
\section{Introduction}%
Quantum phase transitions (QPTs) have always been offering attracting physics~\cite{Vojta2003, Sachdev2011}.
Non-analytic behaviors in the physical observables of an equilibrium interacting system and its universality has been widely explored in various systems:
The spin and fermionic systems such as Ising, Heisenberg, and two-band models~\cite{Chakravarty1989, Schulz1990, Sondhi1997}, interacting bosons~\cite{Kuhner1998, Greiner2002}, and the fermion-boson interaction~\cite{Wang1973, Hwang2015}.
The studies include a large class of one-dimensional (1D) systems in dilute Fermi and Bose gases, the so-called Luttinger liquids, where the problem boils down to a set of bosons with a linear dispersion relation~\cite{Giamarchi2003}. The apparently simplified system exhibits remarkable features~\cite{Gogolin1998} and thanks to the recent technological advances some of their properties have been studied experimentally~\cite{Hofferberth2007, Erne2018, Haller2010, Yang2017}.
The concept of universality in QPTs has also been explored in the nonequilibrium dynamics~\cite{Polkovnikov2011}.
Dynamical behavior of quantum systems far from equilibrium can exhibit abnormalities in the physical observables signaling a dynamical quantum phase transition (DQPT)~\cite{Zvyagin2016, Heyl2018}.
The DQPT is an emerging topic expanding the concept of phase transition to the seemingly uncontrollable parameter, namely \textit{time}, and lies at the heart of sudden quenches~\cite{Sengupta2004, Zurek2005, Kollath2007, Braun2015}.
In many situations a quench across an equilibrium critical point triggers the DQPT~\cite{Schuetzhold2006, Diehl2010, Cazalilla2011, Heyl2013, Huang2016, Halimeh2017, Jurcevic2017, Zunkovic2018, Jafari2019, Lacki2019}.
Nonetheless, the effect is not necessarily associated to the equilibrium state of the system~\cite{Hickey2014, Zauner2017, Jafari2019}.

Bose-Einstein condensates (BEC) at weakly interacting regime are described by a set of non-interacting bosons with a linear dispersion relation~\cite{Recati2005}.
In a very recent experiment the geometric phase of a BEC is measured via a dynamical quantum Zeno effect~\cite{Do2019}.
Such a system can be brought into interaction with an atomic degree of freedom, e.g. via laser transitions. The system has shown to provide interesting physics, especially, it has a close correspondence with the spin-boson model at sufficiently low energy and temperatures~\cite{Fedichev2003, Recati2005, West2013}. When the above mentioned interaction is tuned properly, pure dephasing of the atomic state can be used as a non-destructive probe to the temporal phase fluctuations of the BEC~\cite{Bruderer2006}.

In this paper, we show that despite its seemingly trivial dynamics a system of non-interacting bosons with linear dispersion relation, e.g. a low energy BEC, experiences dynamical phase transitions owing to the quantum interference effects in its topological phase.
The system undergoes a DQPT after a sudden quench out of its equilibrium by a static force. This takes the collective state of the system from a vacuum into a coherent state that accumulate geometric phase over time.
By studying the Loschmidt amplitude, we find that the return rate function behaves non-analytically that is associated with the kinks in the collective geometric phase at the critical times.
The occurrence of a DQPT is evidenced by showing that the cusps occur where the Fisher zeros of the Loschmidt amplitude cross the real time axis, the critical times $t_c$. By investigating the behavior of return amplitude rate function at the critical points we find different scaling behaviors that depend on the spectral density of the static force.

We use our proposed setup for spin--boson model based on color centers in hexagonal boron nitride (h-BN) monolayers~\cite{Abdi2018a} to show such DQPT in the set of bosons manifests itself as kinks in the qubit coherence dynamics.
Such a spin--boson model is realized by coupling the electronic spin of a color center to vibrations of a free-standing h-BN membrane~\cite{Abdi2018}.

%
%
\section{Model}%
According to the Bogoliubov theory, the physics of a BEC at low energies is described by a set of non-interacting bosons with a \textit{linear dispersion relation}:
$\hat H= \sum_{p\neq 0} \upsilon|p|\hatd{a}_{p} \hat{a}_{p}$.
Here, $\upsilon$ is the velocity of sound excitations and $p$ are momenta of the bosons~\cite{Haldane1981, Giamarchi2003}.
The momentum is discrete $p=k\pi/L$ for a finite system length $L$ with integer $k$, and thus, without loss of generality the above Hamiltonian can be written in terms of the discrete modes where frequency of the $k$th mode is given by $\omega_k = 2k\pi \upsilon/L$.
Therefore, the period of the whole system is given by the longest local period $\tau_0 \equiv L/\upsilon$.
Here, we emphasize that the linear dispersion relation is at the heart of the observations that are discussed in this paper. In fact, the DQPT does not occur for systems with a nonlinear dispersion relation.
\begin{figure}[tb]
\includegraphics[width=\columnwidth]{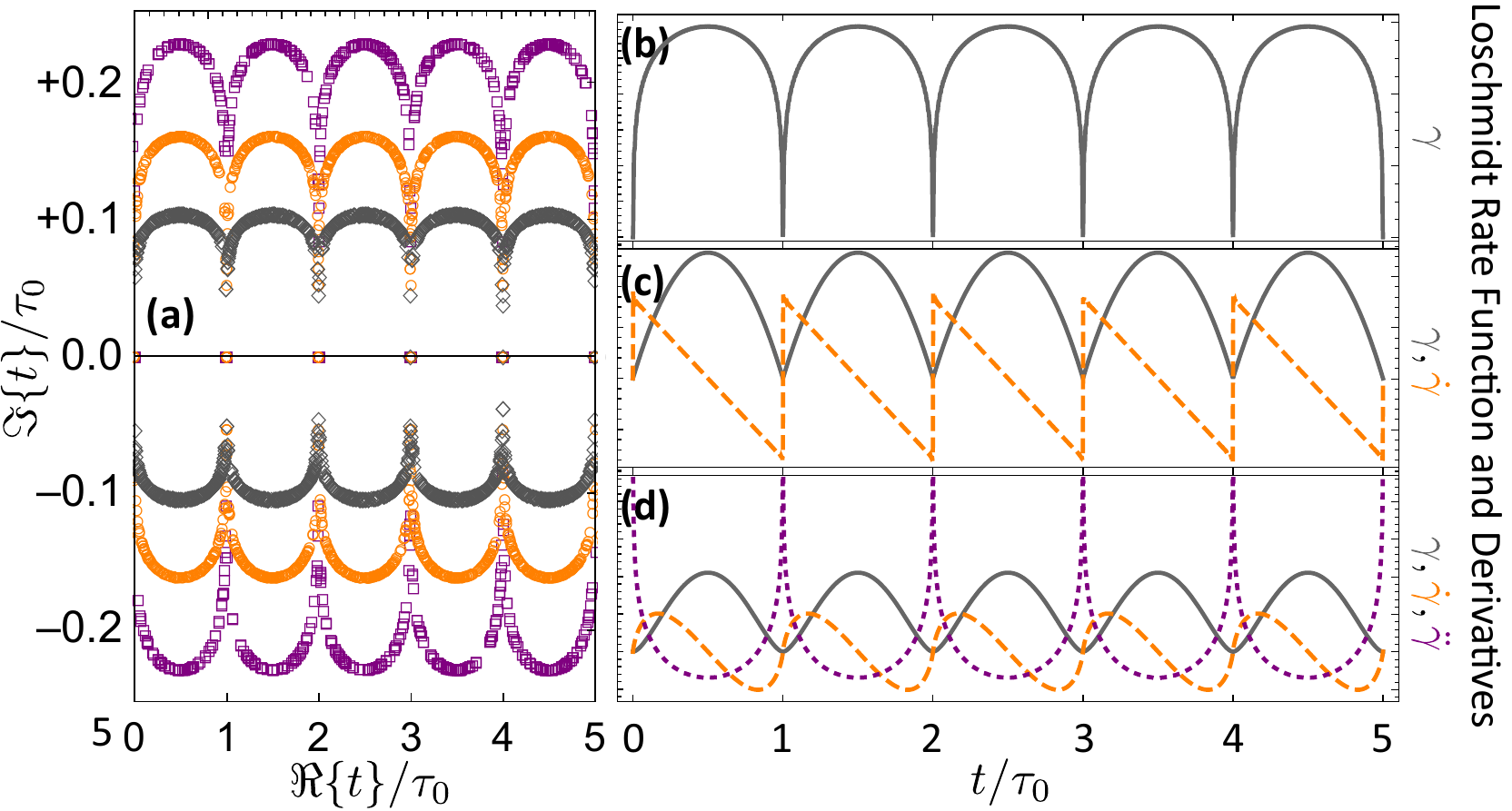}
\caption{%
(a) Fisher zeros for $\alpha=+1$ (gray diamonds), $\alpha=0$ (orange circles), and $\alpha=-1$ (purple squares).
The right panel shows Loschmidt rate function $\gamma$ (gray solid) and its first (dashed orange) and second (dotted purple) derivatives for (b) $\alpha=+1$, (c) $\alpha=0$, and (d) $\alpha=-1$.
The system size is $L=10^3$.
}
\label{fig:DPT}%
\end{figure}

The Hamiltonian of the system thus reads $\hat H_0 = \sum_{k}\omega_k\hatd a_k\hat a_k$.
The ground state of $\hat{H}_0$ is a BEC $\ket{\emptyset}=\bigotimes_k\ket{0}_k$ whose time evolution is trivial and only collects a global dynamical phase corresponding to the vacuum energies.
Nevertheless, a global displacement of the modes can dramatically change the system dynamics as it becomes clear in the following.
By imposing the static force $\vec{\lambda}=\{\lambda_k\}$ on the system, its Hamiltonian turns into $\hat H_1=\hat H_0 +\sum_k \lambda_k(\hat a_k +\hatd a_k)$ whose ground state is a multimode coherent state $\bigotimes_k\ket{\alpha_k}$ with amplitudes $\alpha_k = 2\lambda_k/\omega_k$~\cite{Bennett2012, Asadian2016}. In the phase space representation, each mode is associated to a circular bulb satisfying the minimum Heisenberg uncertainty relation in the respective phase $x_kp_k$ plane. Time evolution of the system with the displacement becomes nontrivial as each bulb goes round a circle in their phase plane passing through the origin and centered at $(x_k,p_k)=(\lambda_k/\omega_k,0)$~\cite{Asadian2014}.
The modular variables of each mode are defined by $\hat{a}_k=(\hat{x}_k+i\hat{p}_k)/\sqrt{2}$ and hence follow the algebra with all of their commutators equal to zero but $[\hat{x}_j,\hat{p}_k]=i\delta_{jk}$.
Due to the non-commuting nature of the modular variables each mode accumulates a geometric phase as it evolves by time, which is proportional to the area enclosed by its circular path: $\phi_k(t)=(\lambda_k/\omega_k)^2[\omega_k t -\sin\omega_k t]$~\cite{Asadian2014}.
The total geometric phase is then
\begin{equation}
\Phi_{\rm G}(t)=\sum_{k}(\frac{\lambda_k}{\omega_k})^2\big[\omega_k t -\sin\omega_k t\big],
\label{geo}
\end{equation}
which is composed of a linear term in time and an oscillating part. For the sake of clarity, we discard the effect of linear term in the following as it does not contribute in the dynamics that we are interested in.
In the cases where $\{\lambda_k\}$ assume \textit{proper} values, the total geometric phase exhibits non-analytic behaviors due to quantum interference effects.
The kinks in the geometric phase, in turn, signal dynamical criticality of the system~\cite{Carollo2005, Zhu2006, Hamma2006} and---as it will become clear shortly---they indeed refer to a \textit{dynamical quantum phase transition}~\cite{Heyl2013, Zunkovic2018}. The concept which is studied in the rest of paper.
It worths mentioning that geometric phase of the ground state is a topological feature of physical systems~\cite{Fujikawa2005}, whose critical behavior across a QPT in the Ising model has established the connection between the two supposedly irrelevant aspects~\cite{Zhu2006, Yuan2007}.

The order of criticality of $\Phi_{\rm G}$ is determined by the displacement parameters in the following way. Assuming $\lambda_k \propto \omega_k^{\alpha/2}$, it is useful to work with $\mathcal{J}(\nu) = \sum_k \lambda_k^2 \delta(\nu -\omega_k)$, the spectral density function. The linear dispersion relation of our system simplifies this function to $\mathcal{J}(\nu) \propto \nu^{\alpha}$, in the continuum limit. 
We find that for $\alpha = 1$ the geometric phase of the system suffers zeroth order criticality at the global periods $\tau_0$. Instead, it experiences a first order criticality (kinks in the first derivative of total geometric phase $\dot\Phi_{\rm G}$) around $\alpha = 0$, and so forth such that for a given integer value of $\alpha$ a $(1-\alpha)$th criticality is expected.
Remarkably, the effect disappears as soon as the dispersion relation deviates from a linear function as shown in Appendix~\ref{sec:geom}.
\begin{figure}[b]
\includegraphics[width=\columnwidth]{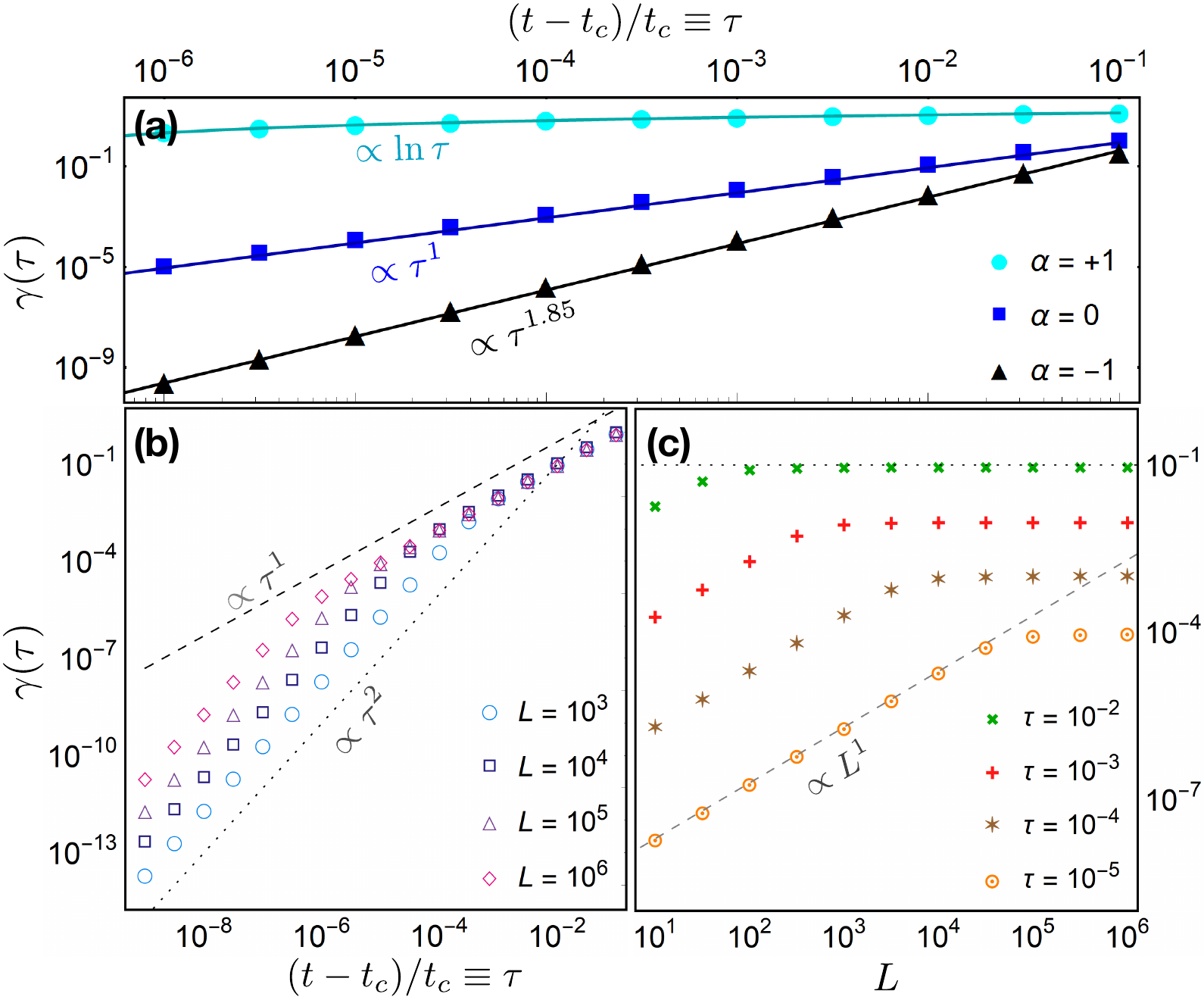}
\caption{%
(a) Behavior and scaling of the return rate function close to the critical times for $\alpha=+1,0,-1$ when the system size is fixed to $L=10^6$.
The power law scaling for $\alpha=0$ is shown in (b) and (c)  with respect to the system size $L$ and normalized critical time proximity $\tau \equiv (t-t_c)/t_c$.
}
\label{fig:scale}%
\end{figure}
%

%
%
\section{Dynamical quantum phase transition}%
In addition to the geometric phase, the critical behavior of the system manifests itself as kinks in the Loschmidt amplitude of the quench dynamics. When the system is prepared in the ground state of $\hat H_0$ and then suddenly pushed by a static force, whose dynamics is dictated by $\hat H_1$, there will be a finite probability for the system to return to the initial state $\ket{\emptyset}$ over time.
The amplitude of such return probabilities, the so-called Loschmidt amplitude, is given by 
\begin{equation}
\mathcal{G}(t)\equiv \bra{\emptyset}\exp\!\{-it\hat{H}_1 \big\}\ket{\emptyset}.
\label{loschmidt}
\end{equation}
This quantity is the dynamical counterpart of the field theoretical boundary partition function and its roots in the complex plane of time, the Fisher zeros, determine nonanalytic properties of the system.
A dynamical phase transition occurs where locus of the zeroes of $\mathcal{G}(z)$ in the complex time plane cross the real axis in the thermodynamic limit~\cite{Heyl2013}.
In Appendix~\ref{sec:losch} we derive an analytical expression for the Loschmidt amplitude in our system $\mathcal{G}(z)=e^{-\gamma(z)}$ with $\gamma(z)\equiv\sum_k(\lambda_k/\omega_k)^2(1-\cos\omega_k z)$, the \textit{return rate function}.
The complex zeros for $\alpha=0, \pm 1$ are then found numerically and one notices that they approach the real time axis at $t_n^*=n\tau_0$ for $n=1,2,3,\cdots$ establishing the occurrence of a DQPT in our model [Fig.~\ref{fig:DPT}(a)].
The transitions are closely related to nonanalytic behavior of the return rate function as well as the total geometric phase of the system.
$\gamma(t)$ remains an analytic function of time except at the critical points $t_c$ that are determined by the Fisher zero lines.
In Fig.~\ref{fig:DPT} this is shown for three different spectral densities with $\alpha = 0, \pm 1$. The order of criticality in the return rate function follows the pattern described above for the total geometric phase. That is, for integer values of $\alpha$ the system experiences a $(1-\alpha)$th DQPT.
Remarkably, as the value of $\alpha$ decreases from $+1$ to $-1$ the density of Fisher zeros around the real time axis decrease and take distance from the axis. This suggests an explanation for the order of DQPT at different $\alpha$ values.

The criticality in the collective dynamics of the system is a quantum interference effect that should be traced back to the geometrical correlations between the modular variables that manifest themselves in the total geometric phase.
Higher order time correlations are anticipated to further unveil the nature of phase transition~\cite{Asadian2014}.

%
%
\subsection*{Scaling}%
We next study scaling of the phase transitions.
To this end, the dynamical behavior of the Loschmidt rate function, which is dynamical counterpart of the free energy function, is studied close to the critical points.
We first notice that the scaling behavior depends on the spectral density exponent $\alpha$. For $\alpha = 1$ the rate function behaves logarithmic in the vicinity of the critical times $t_c=t_n^*$. That is, $\gamma(\tau)\propto \ln\abs{\tau}$ where $\tau\equiv (t-t_c)/t_c$ quantifies proximity to the critical time. Meanwhile, for $\alpha=0$ and $-1$ it has a power law scaling $\gamma(\tau)\propto \abs{\tau}^{\xi_\alpha}$ with dynamical exponents depending on $\alpha$. It is worth mentioning that such behaviors are exactly the same for all critical times.
The results are shown in Fig.~\ref{fig:scale}(a). For a white-noise spectral density we find $\xi_0=1$, while for the pink-noise the exponent is $\xi_{-1}\approx 1.85$.

To further explore the scaling properties of the DQPT, we take the white-noise case and notice that for a given length of the system $L$, the accuracy of the closeness to the critical point $\tau\equiv(t-t_c)/t_c$ is limited to $\tau \gtrsim L^{-1}$. Therefore, the true scaling holds only for the higher values of $\tau$ and for shorter times it reduces to that of leading term of the cosine, i.e. $\propto\tau^2$ [Fig.~\ref{fig:scale}(b)].
On the other hand, for a given value of $\tau$ at the critical points the return rate function scales linearly with the system length up until it reaches the validity limit $L\gtrsim 1/\tau$ from where it turns into a plateau showing insensitivity to the system size close to the critical point [Fig.~\ref{fig:scale}(c)].

\begin{figure}[b]
\includegraphics[width=\columnwidth]{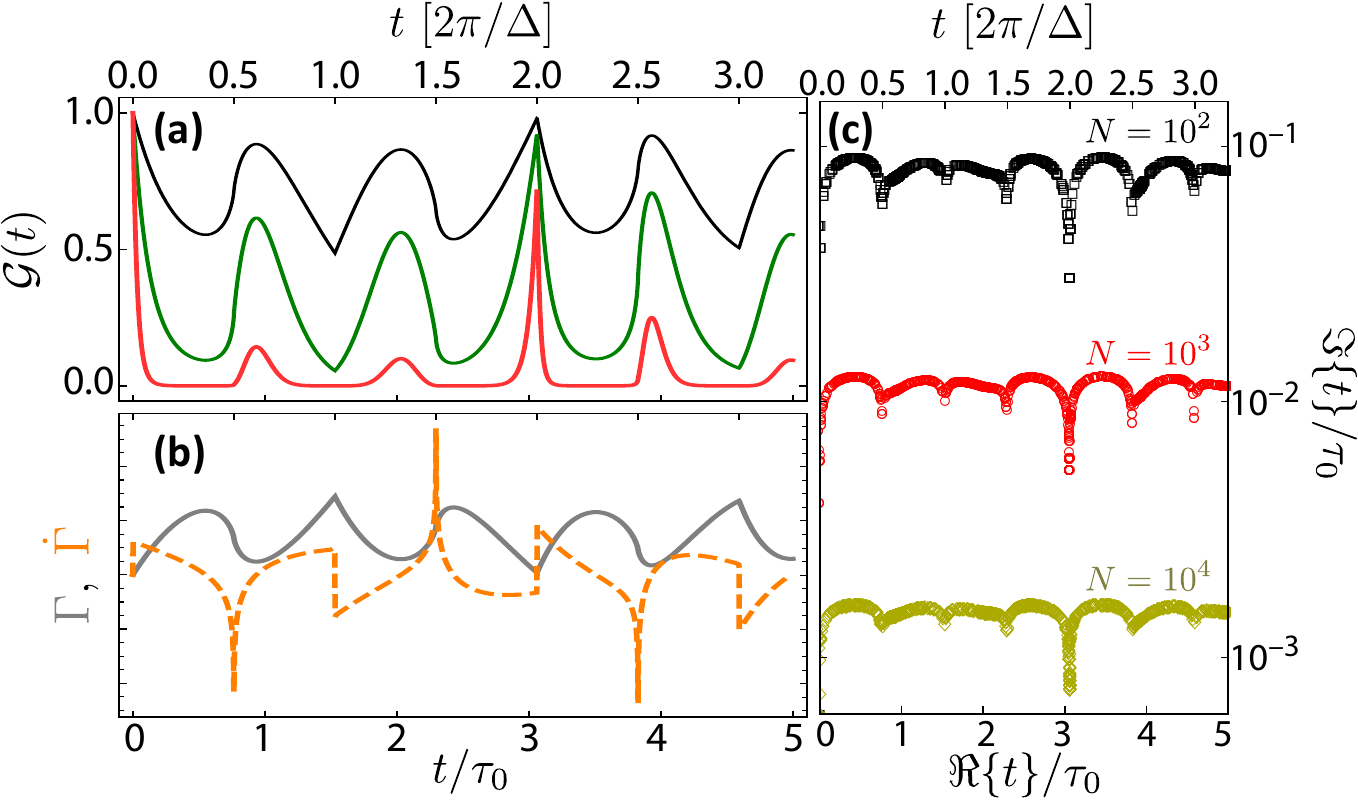}
\caption{%
(a) Collapse and revival in the spin coherence at the coupling rates: $\Lambda_0/\Omega_0=0.5$ (black), $\Lambda_0/\Omega_0=1.0$ (green), and $\Lambda_0/\Omega_0=2.0$ (red).
(b) The return rate function (gray) and its derivative (orange dashed) at zero temperature.
(c) Fisher zeros of the Loschmidt amplitude for three different system sizes. The zeros approach to the real time axis as the system size increases. The tongues suggest that the crossings occur at the integer multiples of $\pi/\Delta$ in the thermodynamical limit.
}
\label{fig:fid}%
\end{figure}
%

%
%
\section{Implementation}%
To experimentally implement a system that shows the above mentioned nontrivial dynamics, here we put forth a setup with a detection scheme.
The system is composed of a free-standing monolayer h-BN membrane with an embedded color center~\cite{Tran2015, Abdi2017, Abdi2018a}.
The latter is responsible for displacement of the modes and also is employed as a probe to the return rate function, and thus, the global geometric phase of the system~\cite{Yuan2007}. The color center has a spin doublet electronic ground state $\{\ket{\uparrow}, \ket{\downarrow} \}$. When immersed in a magnetic field gradient, the spin couples to the position of the membrane $\hat X$. The dynamics is described by the interaction Hamiltonian $\hat H_{\rm int} = \gele\mu_{\rm B}\eta\proj{\uparrow} \hat X$, where $\eta$ is the magnetic field gradient, while $\mu_{\rm B}$ and $\gele$ are electron Bohr magneton and g-factor, respectively~\cite{Abdi2018a}.
For implementing a uniform static force (the white-noise spectral density) on a system of bosons with linear dispersion relation, we take a circular geometry: a membrane with radius $R$ and thickness $h$ with the defect at its center. Therefore, the spin only couples to the axisymmetric modes~\cite{Landau1975, Abdi2016, Abdi2018a}. The membrane is assumed to be subject to a dominant tensile force at the boundaries $\varepsilon \gg (h/R)^2$. The motion of the membrane is then described by its normal vibrational mode spectrum $\{\Omega_k\}$ that span from the lowest frequency $\Omega_0$, the fundamental mode, to the highest $\Omega_N$, set by the maximum vibrational wavelength [Appendix~\ref{sec:elas}].
We expand the mechanical position operator in terms of the normal modes $\hat X=\sum_{k=0}^N x_{k}^{\circ}(\hat b_k +\hatd b_k)$ to arrive at the total Hamiltonian
\begin{equation}
\hat H = \sum_{k=0}^{N}\Omega_k\hatd b_k\hat b_k +\Lambda_k\proj{\uparrow} (b_k +\hatd b_k).
\end{equation}
Here, the bosonic annihilation (creation) operator $\hat b_k$ ($\hatd b_k$) is assigned to the $k$th mechanical normal mode. The coupling strength of each mode to the spin-qubit is given by $\Lambda_k \equiv \gele\mu_{\rm B}\eta x_{k}^{\circ}$ with the zero-point fluctuation amplitude of $k$th mode $x_{k}^{\circ}=\sqrt{\hbar/2M_k^*\Omega_k}$, where $M_k^*$ is the effective mass.
Such vibrational bath is found to have normal modes with equally spaced frequencies $\Omega_{k+1}-\Omega_k = \Delta$, hence they form a linear dispersion relation. And all modes are equally coupled to the spin $\Lambda_k=\Lambda_0$, thus a uniform static force. Therefore, its spectral density is of white-noise with $\alpha=0$~\cite{Abdi2018a}.

\begin{figure}[tb]
\includegraphics[width=\columnwidth]{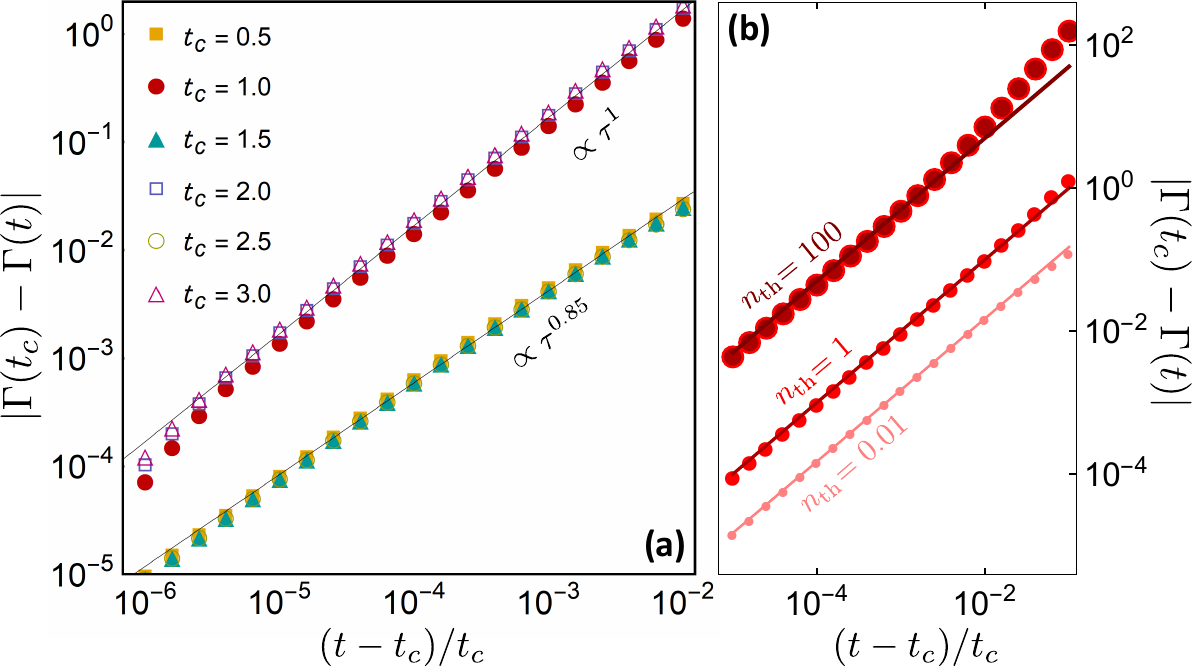}
\caption{%
(a) Scaling of the return rate function close to the critical times (in units of $\overline\tau$). The power law scaling assumes different exponents at the integer and half-integer times where the transitions are first and second order, respectively.
(b) The power law scaling at $t_c=\overline\tau$ for three finite temperatures in units of fundamental mode occupation number $n_{\rm th}$. The lines are linear fits.
In all plots the system size is $N=10^5$.
}
\label{fig:impel}%
\end{figure}
%

%
%
\subsection{Pure dephasing}%
Let us now study dynamics of the qubit when initially prepared in a superposition state $\frac{1}{\sqrt{2}}(\ket{\uparrow}+\ket{\downarrow})$. The off-diagonal elements of the spin density matrix $\varrho$ evolve as
$\bra{\uparrow}\varrho(t)\ket{\downarrow} = \bra{\downarrow}\varrho(t)\ket{\uparrow} = \bra{\uparrow}\varrho(0)\ket{\downarrow}e^{-\Gamma(t)}$
with~\cite{Breuer2007}
\begin{equation}
\Gamma(t)=\sum_{k=0}^N \big(\frac{\Lambda_k}{\Omega_k}\big)^2 \coth\!\big(\frac{\hbar\Omega_k}{2k_{\rm B} T}\big)(1-\cos\Omega_k t),
\label{deph}
\end{equation}
for a thermal bosonic bath at temperature $T$, while $k_{\rm B}$ is the Boltzmann constant.
For a zero-temperature bath $T=0$ it is enough to drop the $\coth$ factor.
One then obviously finds that the free-induction decay (FID) of the spin-qubit is the same as the Loschmidt amplitude of the bosonic bath when quenched from a non-displaced form to a `properly' displaced bath: $\mathcal{G}(t)=\exp\{-\Gamma(t)\}$.
The FID is plotted in Fig.~\ref{fig:fid}(a). The coherence of the qubit falls off with time. However, close to global period of the bath $\tau_0$ it sets for a partial revival depending on the strength of the qubit-bath coupling $\Lambda_0$.
Remarkably, kinks and cusps are observable during the collapse and revivals around half of the collective periods $\overline\tau \equiv 2\pi/\Delta$. As discussed above, such behavior signals a DQPT in the bosonic bath manifesting itself in the coherence dynamics of the spin.
The numerically computed Fisher zeros indeed confirm this observation. One clearly sees from Fig.~\ref{fig:fid}(c) that the Fisher zeros exhibit tongues appearing exactly at the integer multiples of $\overline\tau/2$.
By increasing size of the system the zeros of the Loschmidt amplitude get closer to the real time axis and cross it in the thermodynamic limit.

We next study the DQPT in our system both qualitatively and quantitatively. For this, the rate function $\Gamma$ and its derivative are numerically evaluated and plotted in Fig.~\ref{fig:fid}(b). One notices that kinks at the half integer multiples of $\overline\tau$ are of second order transitions, while at the integer multiples of the period the system undergoes first order transitions.
In order to quantify them, we study the scaling behavior of the cusps close to the critical times.
The results are summarized in Fig.~\ref{fig:scale}; we find that for all critical times the rate function closes to its critical value as a power-law function $\abs{\Gamma(t)-\Gamma(t_c)}\propto\tau^{\Xi}$. The scaling exponent, nonetheless, is different. For the second order (half integers of $\overline\tau$) we find $\Xi\approx 0.85$, while the first order transitions (at integer multiples of $\overline\tau$) scale with $\Xi=1$ [Fig.~\ref{fig:impel}(a)].

%
%
To take into account the most prohibiting experimental limitation, we now study effect of finite bath temperature. In Fig.~\ref{fig:impel}(b) we plot the scaling behavior at $t_c=\overline\tau$ for three different temperatures in units of fundamental mode occupation number $n_{\rm th}=\big[\exp\{\hbar\omega_0/k_{\rm B}T\} -1 \big]^{-1}$.
The exact calculations show that the kinks and cusps in the FID and the return function survive at sufficiently low temperatures. However, as the temperature rises, the kinks start to turn into smooth edges (see Appendix~\ref{sec:deph} for the plots).
In fact, deviations from the power law scaling set off as the temperature rises but it remains detectable even at temperatures as high as $n_{\rm th}\sim 100$.
For a membrane with fundamental frequency $\Omega_0/2\pi = 20$~MHz this occupation number corresponds to an ambient temperature of $T\approx 0.1$~K.

%
%
\begin{acknowledgements}
The author thanks R. Jafari, A. Asadian, and F. Shahbazi for helpful discussions.
The support by STDPO and IUT through SBNHPCC is acknowledged.
\end{acknowledgements}

\appendix
\section{Loschmidt amplitude}\label{sec:losch}
\begin{figure*}[tb]
\includegraphics[width=1.2\columnwidth]{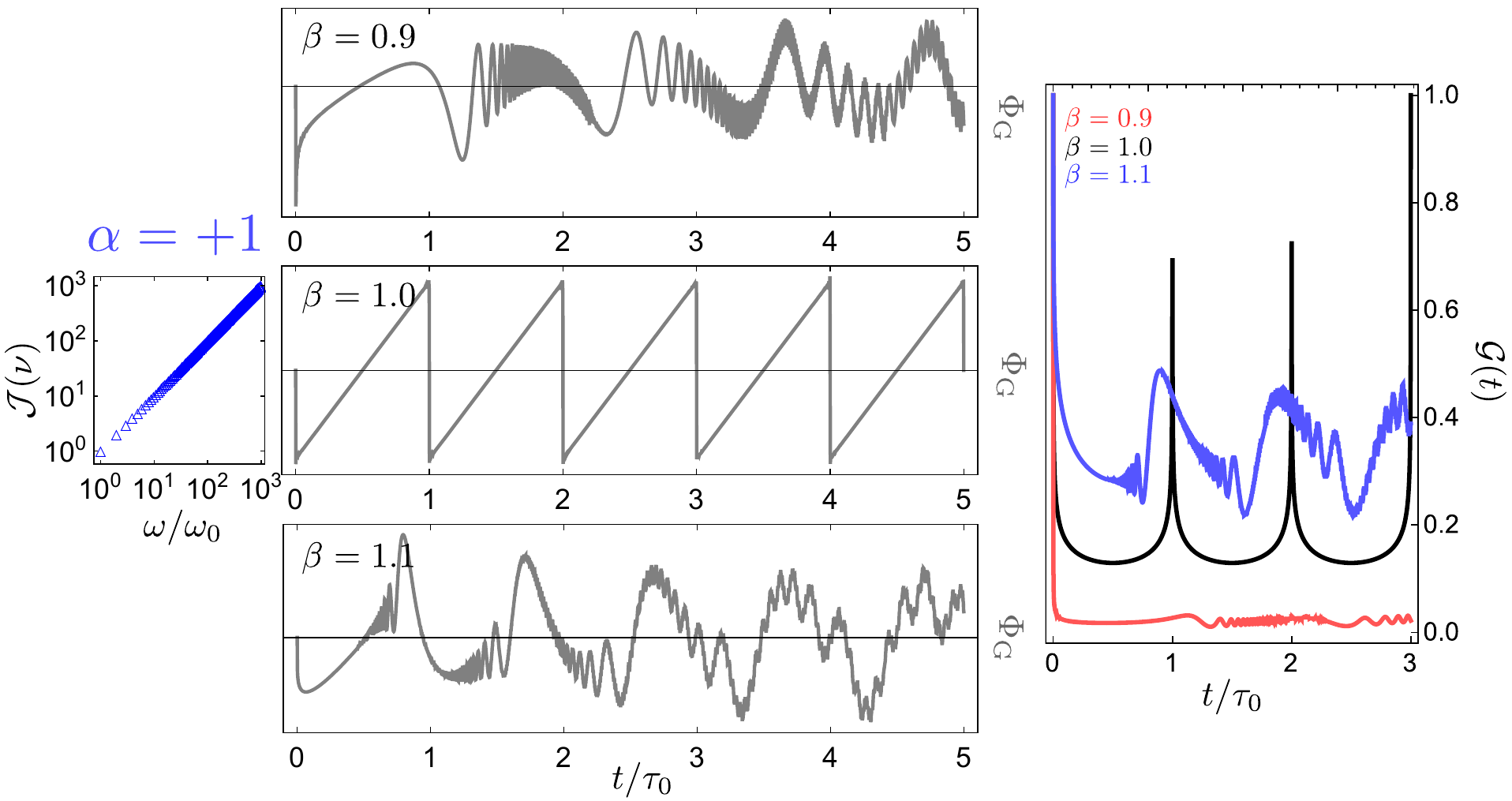} \\
\includegraphics[width=1.2\columnwidth]{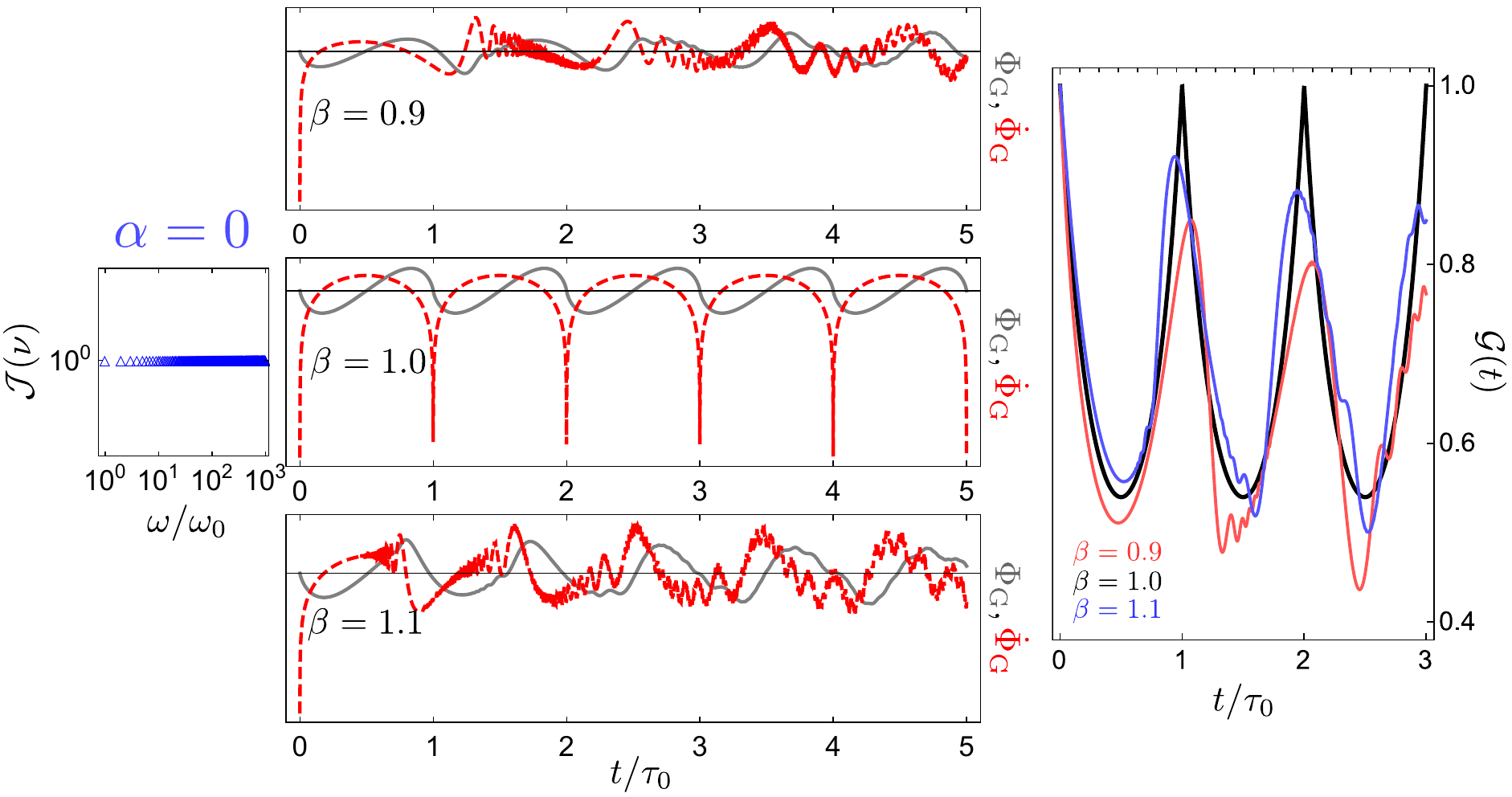} \\
\includegraphics[width=1.2\columnwidth]{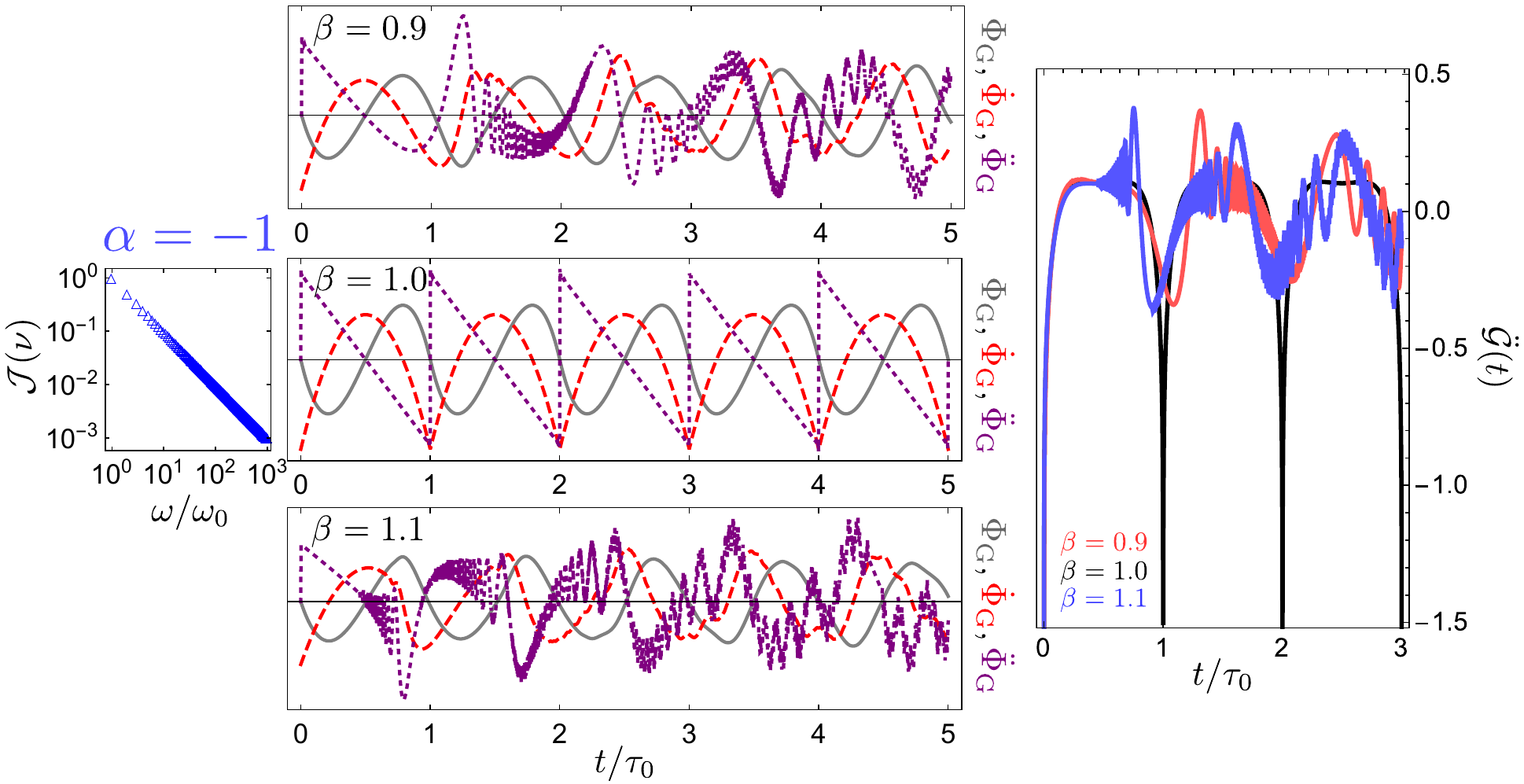}
\caption{%
Topological properties of the model for $\alpha=0,\pm 1$ and at $\beta=0.9,1,1.1$.
The Loschmidt amplitudes are plotted in the right most panels, while the spectral densities are shown at the left panels.
Here, we have taken $N=1000$.
}
\label{fig:GPT}%
\end{figure*}
Loschmidt amplitude is in fact a braket representing the overlap between an initial quantum state and its time evolved version under a different Hamiltonian.
In this appendix we provide the steps of the analytical calculations that lead to the Loschmidt amplitude given in the main text.
In the interaction picture of $\hat H_0$ one arrives at:
\begin{equation*}
e^{-i\hat{H}_0t}\Big(\sum_k\hat{V}_k\Big)e^{i\hat{H}_0t} \equiv \sum_k\hat{V}_k(t) = \sum_k g_k \big(\hatd b_k e^{i\omega_k t} +\hat b_k e^{-i\omega_k t}\big).
\end{equation*}
The interaction picture time evolution is thus given by
\begin{equation}
\hat{U}(t,t_0)=\mathcal{T}_+\exp\Big\{-i\int_{t_0}^t \hspace{-1mm}ds \sum_k\hat{V}_k(s)\Big\},
\label{time}
\end{equation}
where $\mathcal{T}_+$ is the elapsing time ordering operator and $t_0$ is some initial time. The commutator of the $\hat{V}_k(t)$s at two different times are complex numbers:
\begin{equation}
[\hat{V}_k(t),\hat{V}_{k'}(t')]=-2ig_k^2\sin\big[\omega_k(t-t')\big]\delta_{kk'}.
\end{equation}
This property of the interaction Hamiltonian allows us to simplify Eq.~(\ref{time}) into
\begin{align*}
\hat{U}(t,t_0) &=\exp\Big\{i\int_{t_0}^t\hspace{-1.5mm}ds\int_{t_0}^t\hspace{-1.5mm}ds' \Theta(s-s')\varphi(s-s') \Big\} \hat{\mathcal{D}}(t,t_0),
\end{align*}
where $\Theta(t)$ is the Heaviside step function, $\varphi(t-t')\equiv \sum_k g_k^2\sin\omega_k(t-t')$, and $\hat{\mathcal{D}}(t,t_0) \equiv \exp\big\{\sum_k(\alpha_k \hatd b_k -\alpha_k^* \hat b_k) \big\}$ with $\alpha_k=(g_k/\omega_k)(e^{i\omega_k t_0}-e^{i\omega_k t})$.
Therefore, apart from an overall dynamical (time dependent) phase factor the time evolution of the total system is governed by the operator $\hat{\mathcal{D}}(t,t_0)$, which is nothing but a multimode displacement operator.
By assuming an initially ground state for the bath $\ket{\psi_B}(t_0)=\ket{\emptyset}$ the Loschmidt amplitude at every instance of time is
\begin{align}
\mathcal{G}(t)&=e^{i\theta(t,t_0)}\bra{\emptyset}\hat{\mathcal{D}}(t,t_0)\ket{\emptyset}
=e^{i\theta(t,t_0)}\prod_k\bra{\emptyset}\alpha_k\rangle \nonumber\\
&=e^{i\theta(t,t_0)}\prod_k\exp\{-\half|\alpha_k|^2\} \nonumber\\
&= e^{i\theta(t,t_0)}\exp\Big\{-\half\sum_k|\alpha_k|^2\Big\}.
\end{align}
For $t_0=0$ one easily verifies that
\begin{equation}
|\mathcal{G}(t)| = \exp\Big\{-\sum_{k=0}^L \big(\frac{g_k}{\omega_k}\big)^2 (1-\cos\omega_k t)\Big\}.
\end{equation}
This is exactly the equation reported in the paragraph below Eq.~(2) in the main text.

\section{Criticality of the geometric phase}\label{sec:geom}
During the main article our focus has been on the topological order parameter to quantify the criticality in the geometrical properties of the system.
In this appendix, instead, we explore the critical behavior of the geometric phase---which is closely related to the dynamical order parameter introduced in the main text---with more details.
Another purpose of this Appendix is to illustratively emphasize of the central role of the dispersion relation in the geometrical behavior of the bosonic system.
As mentioned before, the critical behaviors studied in the paper are indeed a feature of \textit{linear dispersion relation} of the bath.
To show this illustratively we plot the total geometric phase (sum of the geometric phases accumulated by each boson mode) and its derivatives in Fig.~\ref{fig:GPT}. In the figure the three spectral densities that where studied in the paper are plotted when the dispersion relation is linear and also when it slightly differs from a linear function. Namely, when the bath frequencies are given by $\omega_k = (k+1)^\beta\omega_0$ in terms of the fundamental frequency. Here, we take $\beta = 0.9,~1,~1.1$.
The results clearly verify that the kinks and cusps only appear in the case of linear function, for $\beta=1$. This is evident both from the total geometric phase and the Loschmidt amplitudes that are plotted in the left most panels.
As soon as the dispersion relation deviates from a linear function the non-analytic behaviors turn into rapid oscillations.
The situation will not change even by setting a constant factor to the frequencies. Therefore, the interference of the modal variables through their geometric phase only leads to irregular behavior when the frequency of the modes form a comb-like spectrum. That is, when the frequencies of every two adjacent modes are the same for the whole spectrum.

In Figs.~\ref{fig:GPT} the spectral density function of the three systems with different $\alpha$s (shown above the spectral density plot) studied in the paper are plotted alongside their corresponding geometric phase and the derivatives. The total geometric phase is in gray solid line, its first time derivative is in red dashed line, while the second time derivative is plotted in purple dotted line.
We also plot Loschmidt amplitude in the figure at the different situations described above. In the case of $\alpha=-1$ spectral density the kinks only appear as second order non-analytic points at the critical times. Hence, we show the second derivative of the Loschmidt amplitude $\ddot{\mathcal{G}}(t)$ in this case.

\section{Elasticity}\label{sec:elas}
Let us now present the mechanical model describing the motion of a thin layer of the type we use in the main text. We denote displacement of each point on the membrane by $u(r,\theta)$ with respect to its equilibrium plane. Note that we are assuming a circular membrane with radius $R$. Hence, the polar coordinates are considered. The deflections then obey the following equation of motion \cite{Landau1975},
\begin{equation}
	\rhotd \partial_t^2 u = -D \nabla^2\nabla^2u +T\nabla^2u,
\label{memeqmot}
\end{equation}
where $\nabla^2 u= \frac{1}{r}\frac{\partial}{\partial r}(r\frac{\partial u}{\partial r}) +\frac{1}{r^2}\frac{\partial^2u}{\partial\theta^2}$ is the two-dimensional (2D) Laplacian, $\rhotd$ is the membrane's 2D mass density, and $D = Y h^3/[12(1-\sigma^2)]$ with $Y$ being the Young modulus, $h$ the thickness of the membrane, and $\sigma$ its Poisson ratio. Here, $T = T_0 +\Delta T$ is the total tensile force experienced by the membrane, with $T_0$ the force applied by the support and $\Delta T$ the bending tension as a result of extension. The latter is proportional to the relative change in the surface area of the membrane,
\begin{equation}
	\Delta T = Yh\frac{\Delta A}{A_0} = \frac{Y h}{2A_0}\int_0^{2\pi}\hspace{-2mm}d\theta\int_0^R \hspace{-2mm}rdr\left|\nabla u\right|^2.
\label{bendten}
\end{equation}
We assume that the bending energy is much smaller than the contribution coming from the tensile force on the edges of the membrane ($D\nabla^2\nabla^2u \ll  T_0 \nabla^2u$). Thus the equation (\ref{memeqmot}) simplifies to 
\begin{equation}
	\rhotd \partial_t^2 u = T_0\nabla^2 u +\left(\frac{Y h}{2A_0}\int_0^{2\pi}\hspace{-2mm}d\theta\int_0^R \hspace{-2mm}rdr\left|\nabla u\right|^2\right)\nabla^2 u,
\label{effmemeqmot}
\end{equation}
and leads to nonlinearities in the equation of motion, which can be treated perturbatively in our regime of interest. In particular, we proceed by finding the normal modes associated to the linear problem $\rhotd \partial_t^2 u = T_0 \nabla^2 u$, expand equation (\ref{effmemeqmot}) in these, and keep leading nonlinear terms only. 
The membrane geometry is taken to be a disc with radius $R$, which possess radially-symmetric (axisymmetric) modes as well as excited modes with trigonometric polar-angle dependence.
We apply the separation of variables $u(r,\theta,t) = \sum_{n=0}^\infty X_n(t)\psi_n(r,\theta)$ to obtain the following normal-mode equations for the linear part of Eq.~(\ref{effmemeqmot}),
\begin{equation}
	\nabla^2\psi_n(r,\theta) +\left(\frac{\zeta_n}{R}\right)^2\psi_n(r,\theta) =0,
\label{modes}
\end{equation}
where $\zeta_n^2 = \rhotd R^2\Omega_n^2/T_0$ and should be found numerically as shall be explained below. Here, $\Omega_n\in\mathbb{R}$ is the oscillation frequency of $n$th mode. The nonsingular solutions of the above equation can be written as $\psi_n(r) = \mathcal A_n J_{k_n}(\zeta_n r/R) \cos k_n\theta$, where $J_k$ are Bessel functions of the first kind, $k_n\in\mathbb{Z}$. We order the solutions with increasing values of their separation constant, so that $\zeta_0 < \zeta_1 < \zeta_2 < \cdots $, i.e. the normal modes $\{\psi_n\}_{n=0,1,2,\dots}$ are ordered from lower to higher frequency. For our clamped circular membrane, the boundary conditions imply
\begin{equation}
	J_{k_n}(\zeta_n) =0.
\end{equation}
We choose $\mathcal A_n$ such that $\max\{\psi_n\}=1$ so that $X_n$ provides the maximal vertical deflection of the membrane when the corresponding mode is excited. For the axisymmetric modes $\mathcal{A}_n=1$.

The dynamics of each mode is obtained by inserting $u(r,\theta,t)=\sum_{n=0}^\infty X_n(t)\psi_n(r,\theta)$ into Eq.~(\ref{effmemeqmot}), multiplying both sides of the resulting equation by $\psi_m(r,\theta)$, and integrating over the surface of the membrane. This leads us to
\begin{equation}
	M_n^*\ddot{X}_n = -\mathcal K_{n}X_n -\frac{Y h}{2\pi R^2}\sum_{m} \mathcal M_{mm}\mathcal M_{nn} X_m^2X_n.
\end{equation}
Here, $M_n^* = \rhotd \int_0^{2\pi}\hspace{-1mm}d\theta\int_0^R \hspace{-0.8mm}rdr \psi_n^2$ is the effective mass of the $n$th mode and $\mathcal K_{n} = T_0 \mathcal M_{nn}$ the mode's spring constant, and we have introduced $\mathcal M_{mn} = \int_0^{2\pi}\hspace{-1mm}d\theta\int_0^R \hspace{-0.8mm}rdr\psi_m \nabla^2\psi_n = \int_0^{2\pi}\hspace{-1mm}d\theta\int_0^R \hspace{-0.8mm}rdr(\nabla\psi_m)\cdot(\nabla\psi_n)$. We note that $\mathcal M_{mn}$ vanishes for $m \neq n$ as a result of the orthogonality of the modes, which simplifies the nonlinear cross-coupling between the modes to only two modes (instead of four-wave mixing that would appear for the $T_0=0$ case).
Therefore, the dynamics of the membrane can be described by the Hamiltonian
\begin{align}
H_{\rm m} &= \sum_{n=0}^\infty\Big(\frac{P_n^2}{2M_n^*} +\frac{1}{2}M_n^*\Omega_n^2 X_n^2 +\frac{1}{4}\beta_n X_n^4 \Big) \nonumber\\
&+\frac{Y h}{2\pi R^2} \sum_{n\neq m}^\infty \mathcal M_{nn}\mathcal M_{mm} X_n^2 X_m^2. \label{MultimodeH1}
\end{align}
where the terms with $n=m$ are excluded from the second sum as they are already included in the first sum. Moreover, we have defined $\Omega_n=\sqrt{\mathcal{K}_n/M_n^*}$ and $\beta_n=4(Y h/2\pi R^2)\mathcal M_{nn}^2$.

\begin{figure}[tb]
\includegraphics[width=0.7\columnwidth]{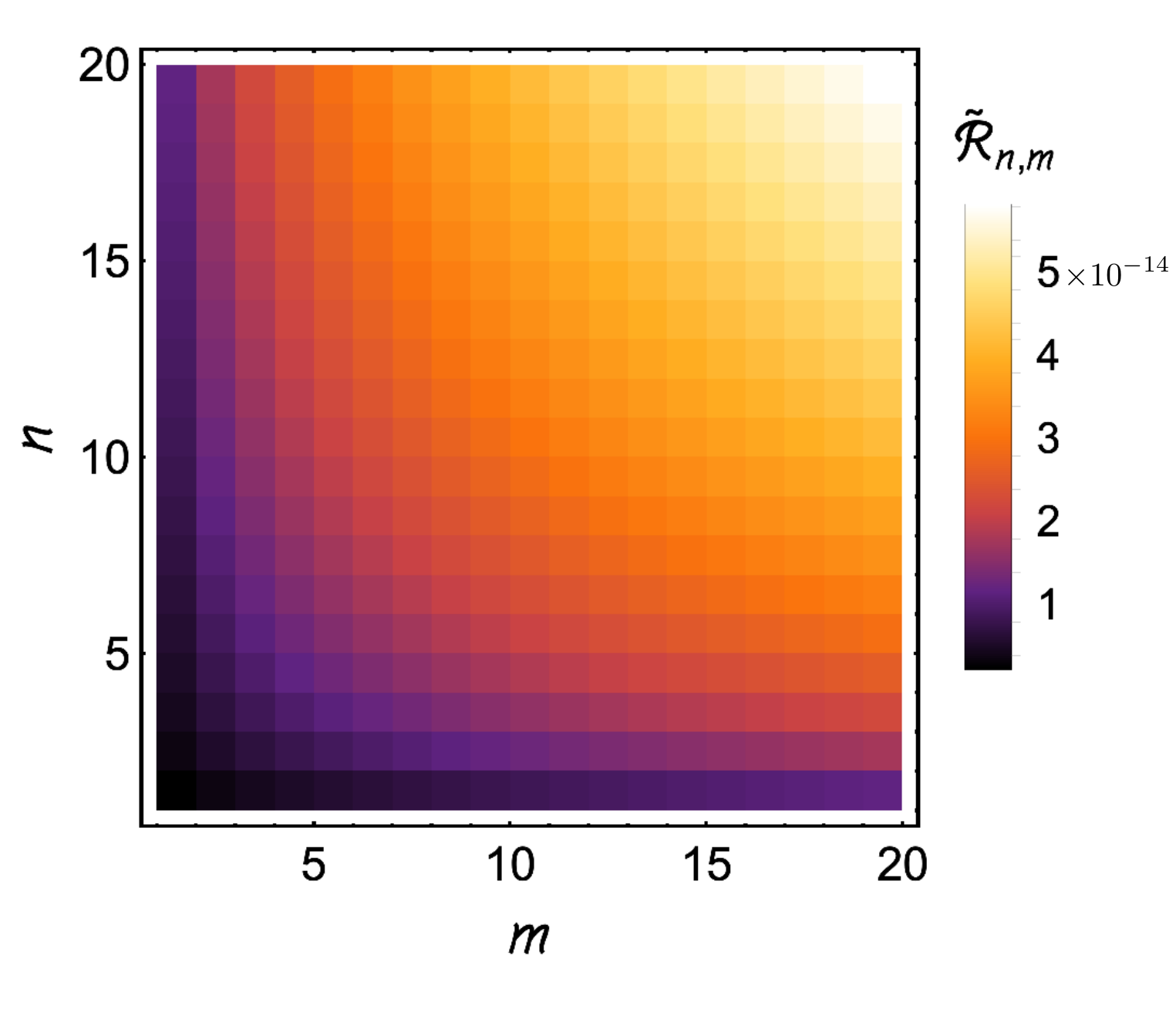}
\caption{%
Density plot of the ratio parameter with respect to the mode numbers for the first 20 modes. One notices that the parameter is maximized for $m=n$.
}
\label{fig:ratio}%
\end{figure}

Let us now put the focus on the axisymmetric modes where the diagonal nodes are absent from the profile. Indeed, spin state of a defect at the center of a circular membrane only couples to the axisymmetric modes~\cite{Abdi2018a}. By this choice, the normal modes are fully identified by the number of their radial nodes and their profile function simplifies to: $\psi_n(r)=J_0(\zeta_n\frac{r}{R})$ where the $\zeta_n$s are determined through $J_0(\zeta_n)=0$.
It is then straightforward to find that $\Omega_n=(\zeta_n/R)\sqrt{Yh\varepsilon/\rhotd}$ and given the following identity for the Bessel functions
\begin{equation}
\int_0^1 x J_{\nu}(\zeta_{\nu,n}x)J_{\nu}(\zeta_{\nu,m}x)dx = \frac{1}{2}\big[J_{\nu+1}(\zeta_{\nu ,n})\big]^2\delta_{mn}~,
\end{equation}
one finds $M_n^*=\pi R^2 \rhotd \big[J_1(\zeta_{n})\big]^2$ for the effective mass of the modes. Furthermore, the Duffing nonlinearity and the inter-mode coupling rates are respectively given by
\begin{equation}
\chi_{nm} = \frac{Yh\pi}{2R^2}\zeta_n^2\zeta_m^2 \big[J_1(\zeta_n)\big]^2\big[J_1(\zeta_m)\big]^2,
\end{equation}
and $\frac{1}{4}\beta_n\equiv \chi_{nn}.$
To compare the harmonic part of the Hamiltonian against the anharmonicity and the quartic inter-mode coupling rates one has to take into account the vibrational amplitudes. In this work we mainly consider temperatures close to the absolute zero as the operating point of the system. Therefore, one takes the thermal amplitudes for the expectation values $\langle X_n\rangle \approx \overline{n}_{n} x_n^\circ$ in a mean-field approximation to arrive at:
\begin{equation}
\mathcal{R}_n\equiv\frac{\frac{1}{4}\beta (\overline{n}_{n}x_{n}^{\circ})^4}{\frac{1}{2}M_n^*\Omega_n^2 (\overline{n}_{n}x_{n}^{\circ})^2} = \frac{\hbar\zeta_n}{8\pi R^3} \frac{\varepsilon^{-\frac{3}{2}}}{\sqrt{Yh\rhotd}}\overline{n}_{n}^2,
\end{equation}
where $\overline{n}_{n}=\big[\exp\{\hbar\Omega_n/k_{\rm B}T\}-1\big]^{-1}$ is the thermal occupation number of the $n$th mode.
We have taken the high tensile strain at the boundaries as our working regime. Therefore, we set $\varepsilon = 10^7\times(h/R)^2$ according to the results presented in Ref.~\cite{Abdi2018a}. By putting the values corresponding to a monolayer of hexagonal boron nitride that are $h=3.3$~\AA, $Y = 270$~GPa, and $\rhotd = 6.93\times 10^{-7}$~kg/m$^2$ we are brought to
\begin{equation}
\mathcal{R}_n \approx 4.69 \times 10^{-16} \zeta_n \overline{n}_{n}^2.
\label{neglig}
\end{equation}
The value of $\zeta_n$, the Bessel function zeros, increases only linearly with the node number, i.e. $\zeta_n \propto n$.
Meanwhile, the occupation number falls rapidly with the increasing frequency for a given temperature.
Therefore, the ratio $\mathcal{R}_n$ remains negligible for a large number of vibrational modes at the temperatures discussed in the main text. Negligibility of the inter-mode coupling is concluded in the same lines. To show this, we define a generalized ratio parameter as the following
\begin{equation}
\widetilde{\mathcal{R}}_{n,m}\equiv \frac{\chi_{nm}\overline{n}_{n}x_{n}^{\circ}\overline{n}_{m}x_{m}^{\circ}}{\frac{1}{2}\sqrt{M_n^*M_m^*}\Omega_n\Omega_m}.
\end{equation}
Note that $\widetilde{\mathcal{R}}_{n,n}=\mathcal{R}_n$. This parameter is plotted in Fig.~\ref{fig:ratio} with the hBN parameter values. One obviously finds that the parameter is maximized at for $n=m$, which a small in our setup as evidenced in Eq.~\eqref{neglig}.
\begin{figure*}[tb]
\includegraphics[width=1.1\columnwidth]{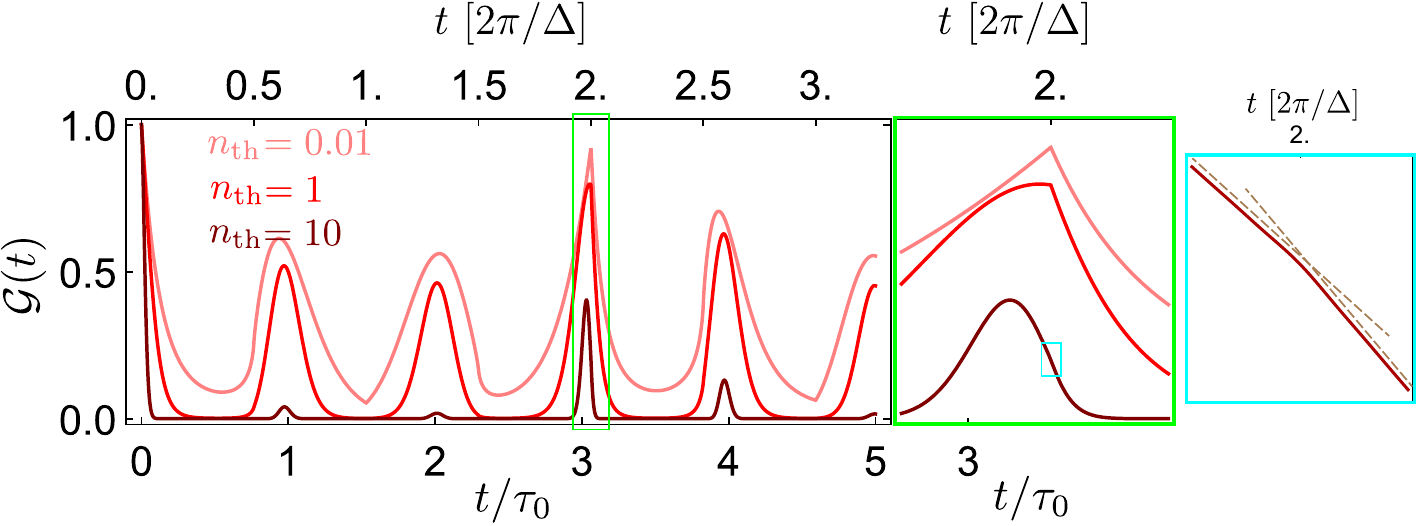}
\caption{%
Collapse and revival in the spin coherence at the coupling rate $g_0/\omega_0=1$ at three different temperatures. The middle panel is a close up of the most prominent cusp that happens at $t_c=2\overline\tau$. The right most panel gives yet a closer look at the high temperature FID illustrating its non-analytic essence at the critical time.
Here, we have taken $N=10^4$.
}
\label{fig:fidtemp}%
\end{figure*}
%

\section{Pure dephasing at finite temperature}\label{sec:deph}
In this section we present the temporal behavior of the system when the temperature is not exactly zero.
The steps of the analytical calculations are quite similar to that of Loschmidt amplitude given in Appendix~A.
The total Hamiltonian in Schr{\"o}dinger picture is $\hat{H}=\hat{H}_0+\hat{H}_{\rm int} $with:
\begin{align}
&\hat{H}_0 = \sum_k\omega_k\hatd{b}_k\hat{b}_k, \\
&\hat{H}_{\rm int} =\ket{\uparrow}\!\bra{\uparrow}\sum_k g_k(\hat{b}_k+\hatd{b}_k).
\end{align}
In the interaction picture of $\hat H_0$ the Hamiltonian becomes time dependent and given by
\begin{equation*}
e^{-i\hat{H}_0t}\hat{H}_{\rm int}e^{i\hat{H}_0t} \equiv \hat{H}_{\rm int}(t) = \ket{\uparrow}\!\bra{\uparrow}\sum_k g_k \big(\hatd b_k e^{i\omega_k t} +\hat b_k e^{-i\omega_k t}\big).
\end{equation*}
The interaction picture time evolution is given by
\begin{equation}
U_{\rm int}(t)=\mathcal{T}_+\exp\Big\{-i\int_{0}^t \hspace{-1mm}ds\ \hat{H}_{\rm int}(s)\Big\},
\label{timedeph}
\end{equation}
The commutator of the $\hat{H}_{\rm int}(t)$ at two different times is a complex number:
\begin{equation*}
[\hat{H}_{\rm int}(t),\hat{H}_{\rm int}(t')]=-2i\sum_k g_k^2\sin\!\big[\omega_k(t-t')\big]\equiv -2i\varphi(t-t').
\end{equation*}
This property of the interaction Hamiltonian allows us to simplify Eq.~(\ref{timedeph}) into
\begin{align}
U_{\rm int}(t) &=\exp\Big\{i\int_{0}^t\hspace{-1.5mm}ds\int_{0}^t\hspace{-1.5mm}ds' \Theta(s-s')\varphi(s-s') \Big\} \hat{\mathcal{V}}(t),
\end{align}
where $\hat{\mathcal{V}}(t) \equiv \exp\big\{\proj{\uparrow}\sum_k(\alpha_k \hatd b_k -\alpha_k^* \hat b_k) \big\}$ with $\alpha_k=(g_k/\omega_k)(1-e^{i\omega_k t})$.
%
\begin{figure*}[tb]
\includegraphics[width=1.2\columnwidth]{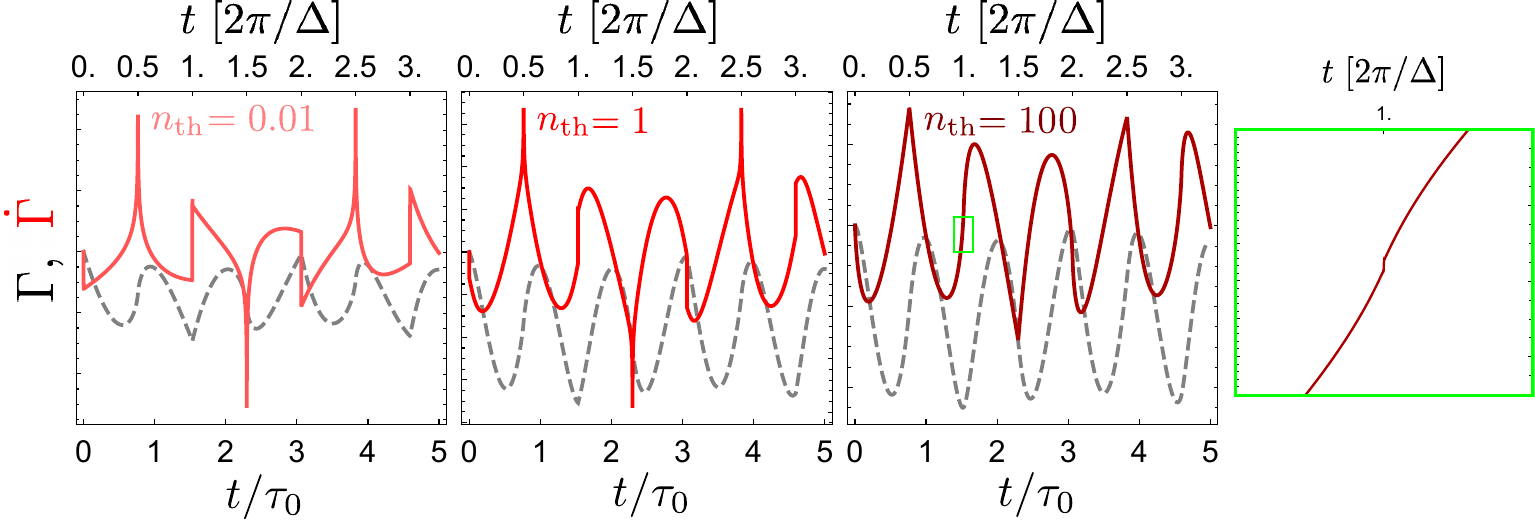}
\caption{%
The return rate function (dashed gray) and its derivative (red varieties) at different temperatures. The right most panel shows a closer look at the jump in $\dot{\nu}_D$ at $t_c=\overline\tau$.
Here, we have taken $N=10^4$.
}
\label{fig:dtptemp}%
\end{figure*}
Therefore, apart from an overall dynamical (time dependent) phase factor the time evolution of the total system is governed by the operator $\hat{\mathcal{V}}(t)$. In fact, $\hat{\mathcal{V}}(t)$ is a state-dependent multimode displacement operator such that:
\begin{align*}
\hat{\mathcal{V}}(t)(\ket\downarrow \otimes \ket\phi) &= \ket\downarrow \otimes \ket\phi, \\
\hat{\mathcal{V}}(t)(\ket\uparrow \otimes \ket\phi) &= \ket\uparrow \otimes \hat{\mathcal{D}}(t,0)\ket\phi.
\end{align*}
We assume an initially separable qubit-bath state
\begin{equation}
R(0)=\varrho(0)\otimes\rho_{B}.
\end{equation}
The matrix elements of the qubit system is then determined by
\begin{equation}
\varrho_{ij}(t)=\bra{i}\varrho(t)\ket{j}=\bra{i}\tr{B}{\mathcal{V}(t)R(0)\mathcal{V}^{-1}(t)}\ket{j},
\end{equation}
where $i,j=\downarrow,\uparrow$. It is easy to verify that the populations stay constant with time. That is, $\varrho_{\downarrow\downarrow}(t)=\varrho_{\downarrow\downarrow}(0)$ and $\varrho_{\uparrow\uparrow}(t)=\varrho_{\uparrow\uparrow}(0)$. While the off-diagonal elements of the density matrix, the coherences, behave as
\begin{equation}
\varrho_{\uparrow\downarrow}(t)=\varrho_{\downarrow\uparrow}^*(t)=\varrho_{\uparrow\downarrow}(0)e^{\Gamma(t)},
\end{equation}
where $\Gamma(t)$, the decoherence function, is calculated from the following equation
\begin{align*}
\Gamma(t) &= \log\bigg(\tr{B}{\exp\Big[\sum_k(\alpha_k\hat{b}_k -\alpha_k^*\hatd{b}_k) \Big]\rho_B}\bigg) \\
&= \sum_k\log\Big(\tr{B}{\exp\big[\alpha_k\hat{b}_k -\alpha_k^*\hatd{b}_k) \big]\rho_B}\Big).
\end{align*}
The expectation value appearing the argument of log functions is the Wigner characteristic function of the bath mode $k$:
\begin{equation}
\chi_{_{\rm W}}(\alpha_k,\alpha_k^*)=\tr{B}{\exp\big[\alpha_k\hat{b}_k -\alpha_k^*\hatd{b}_k) \big]\rho_B}.
\end{equation}
When the state of the bath is Gaussian, e.g. a thermal state or when the bath is in its ground state, this characteristic function is easily determined by noting that:
\begin{equation}
\chi_{_{\rm W}}(\alpha_k,\alpha_k^*)=\exp\bigg[-\half|\alpha_k|^2\tr{B}{\big(\hat{b}_k\hatd{b}_k +\hatd{b}_k\hat{b}_k\big)\rho_B} \bigg].
\end{equation}
Having this and the relation of $\alpha_k$ with system parameters at each instance of time one calculates
\begin{equation}
\chi_{_{\rm W}}(\alpha_k,\alpha_k^*)=-\big(\frac{g_k}{\omega_k}\big)^2 \coth\big(\frac{\hbar\omega_k}{2k_{\rm B} T}\big)(1-\cos\omega_k t).
\end{equation}
Therefore, the decoherence function when the bosonic bath is in a thermal state turns into
\begin{equation}
\Gamma(t)=-\sum_{k=0}^N \big(\frac{g_k}{\omega_k}\big)^2 \coth\big(\frac{\hbar\omega_k}{2k_{\rm B} T}\big)(1-\cos\omega_k t).
\end{equation}
This is the result used in the main text. The scaling behavior is already presented and discussed in the main text. In order to provide supplementary information, here we plot the FID evolution with time at different bath temperatures. The spin-qubit is initially prepared in a superposition state (a maximally coherent state) $(\ket\downarrow +\ket\uparrow)/\sqrt{2}$ with the density matrix:
\begin{equation}
\varrho(0) = \frac{1}{2}\left(
\begin{array}{cc}
	1 & 1 \\
	1 & 1 \\
\end{array}
\right).
\end{equation}
The spin coherence at every instance of time is then given by $\varrho_{\uparrow\downarrow}(t)=\half\mathcal{G}(t)$.
In Fig.~\ref{fig:fidtemp}, $\mathcal{G}(t)$ is plotted for a few periods of the fundamental mode at three different temperatures in units of the occupation number of the fundamental mode $n_{\rm th}$.
One clearly notices that the kinks start to fade out as the temperature rises. Nevertheless, the cusps survive even at temperatures as high as $n_{\rm th}\approx 100$ as the closer looks to the plots show.
To further study effect of finite temperature, we also plot the return rate function and its derivative in Fig.~\ref{fig:dtptemp}. The cusps are still appreciable at low temperatures. However, they are not visually observable at higher temperatures unless in a close-up picture (the right most panel in the figure).

%
%
\bibliography{DTPT}

\begin{thebibliography}{55}%
\makeatletter
\providecommand \@ifxundefined [1]{%
 \@ifx{#1\undefined}
}%
\providecommand \@ifnum [1]{%
 \ifnum #1\expandafter \@firstoftwo
 \else \expandafter \@secondoftwo
 \fi
}%
\providecommand \@ifx [1]{%
 \ifx #1\expandafter \@firstoftwo
 \else \expandafter \@secondoftwo
 \fi
}%
\providecommand \natexlab [1]{#1}%
\providecommand \enquote  [1]{``#1''}%
\providecommand \bibnamefont  [1]{#1}%
\providecommand \bibfnamefont [1]{#1}%
\providecommand \citenamefont [1]{#1}%
\providecommand \href@noop [0]{\@secondoftwo}%
\providecommand \href [0]{\begingroup \@sanitize@url \@href}%
\providecommand \@href[1]{\@@startlink{#1}\@@href}%
\providecommand \@@href[1]{\endgroup#1\@@endlink}%
\providecommand \@sanitize@url [0]{\catcode `\\12\catcode `\$12\catcode
  `\&12\catcode `\#12\catcode `\^12\catcode `\_12\catcode `\%12\relax}%
\providecommand \@@startlink[1]{}%
\providecommand \@@endlink[0]{}%
\providecommand \url  [0]{\begingroup\@sanitize@url \@url }%
\providecommand \@url [1]{\endgroup\@href {#1}{\urlprefix }}%
\providecommand \urlprefix  [0]{URL }%
\providecommand \Eprint [0]{\href }%
\providecommand \doibase [0]{https://doi.org/}%
\providecommand \selectlanguage [0]{\@gobble}%
\providecommand \bibinfo  [0]{\@secondoftwo}%
\providecommand \bibfield  [0]{\@secondoftwo}%
\providecommand \translation [1]{[#1]}%
\providecommand \BibitemOpen [0]{}%
\providecommand \bibitemStop [0]{}%
\providecommand \bibitemNoStop [0]{.\EOS\space}%
\providecommand \EOS [0]{\spacefactor3000\relax}%
\providecommand \BibitemShut  [1]{\csname bibitem#1\endcsname}%
\let\auto@bib@innerbib\@empty
\bibitem [{\citenamefont {Vojta}(2003)}]{Vojta2003}%
  \BibitemOpen
  \bibfield  {author} {\bibinfo {author} {\bibfnamefont {M.}~\bibnamefont
  {Vojta}},\ }\bibfield  {title} {\bibinfo {title} {Quantum phase
  transitions},\ }\href {https://doi.org/10.1088/0034-4885/66/12/R01}
  {\bibfield  {journal} {\bibinfo  {journal} {Rep. Prog. Phys.}\ }\textbf
  {\bibinfo {volume} {66}},\ \bibinfo {pages} {2069} (\bibinfo {year}
  {2003})}\BibitemShut {NoStop}%
\bibitem [{\citenamefont {Sachdev}(2011)}]{Sachdev2011}%
  \BibitemOpen
  \bibfield  {author} {\bibinfo {author} {\bibfnamefont {S.}~\bibnamefont
  {Sachdev}},\ }\href@noop {} {\emph {\bibinfo {title} {Quantum Phase
  Transitions}}},\ \bibinfo {edition} {2nd}\ ed.\ (\bibinfo  {publisher}
  {Cambridge University Press},\ \bibinfo {address} {New York},\ \bibinfo
  {year} {2011})\BibitemShut {NoStop}%
\bibitem [{\citenamefont {Chakravarty}\ \emph {et~al.}(1989)\citenamefont
  {Chakravarty}, \citenamefont {Halperin},\ and\ \citenamefont
  {Nelson}}]{Chakravarty1989}%
  \BibitemOpen
  \bibfield  {author} {\bibinfo {author} {\bibfnamefont {S.}~\bibnamefont
  {Chakravarty}}, \bibinfo {author} {\bibfnamefont {B.~I.}\ \bibnamefont
  {Halperin}},\ and\ \bibinfo {author} {\bibfnamefont {D.~R.}\ \bibnamefont
  {Nelson}},\ }\bibfield  {title} {\bibinfo {title} {Two-dimensional quantum
  heisenberg antiferromagnet at low temperatures},\ }\href
  {https://doi.org/10.1103/PhysRevB.39.2344} {\bibfield  {journal} {\bibinfo
  {journal} {Phys. Rev. B}\ }\textbf {\bibinfo {volume} {39}},\ \bibinfo
  {pages} {2344} (\bibinfo {year} {1989})}\BibitemShut {NoStop}%
\bibitem [{\citenamefont {Schulz}(1990)}]{Schulz1990}%
  \BibitemOpen
  \bibfield  {author} {\bibinfo {author} {\bibfnamefont {H.~J.}\ \bibnamefont
  {Schulz}},\ }\bibfield  {title} {\bibinfo {title} {Incommensurate
  antiferromagnetism in the two-dimensional hubbard model},\ }\href
  {https://doi.org/10.1103/PhysRevLett.64.1445} {\bibfield  {journal} {\bibinfo
   {journal} {Phys. Rev. Lett.}\ }\textbf {\bibinfo {volume} {64}},\ \bibinfo
  {pages} {1445} (\bibinfo {year} {1990})}\BibitemShut {NoStop}%
\bibitem [{\citenamefont {Sondhi}\ \emph {et~al.}(1997)\citenamefont {Sondhi},
  \citenamefont {Girvin}, \citenamefont {Carini},\ and\ \citenamefont
  {Shahar}}]{Sondhi1997}%
  \BibitemOpen
  \bibfield  {author} {\bibinfo {author} {\bibfnamefont {S.~L.}\ \bibnamefont
  {Sondhi}}, \bibinfo {author} {\bibfnamefont {S.~M.}\ \bibnamefont {Girvin}},
  \bibinfo {author} {\bibfnamefont {J.~P.}\ \bibnamefont {Carini}},\ and\
  \bibinfo {author} {\bibfnamefont {D.}~\bibnamefont {Shahar}},\ }\bibfield
  {title} {\bibinfo {title} {Continuous quantum phase transitions},\ }\href
  {https://doi.org/10.1103/RevModPhys.69.315} {\bibfield  {journal} {\bibinfo
  {journal} {Rev. Mod. Phys.}\ }\textbf {\bibinfo {volume} {69}},\ \bibinfo
  {pages} {315} (\bibinfo {year} {1997})}\BibitemShut {NoStop}%
\bibitem [{\citenamefont {K{\"u}hner}\ and\ \citenamefont
  {Monien}(1998)}]{Kuhner1998}%
  \BibitemOpen
  \bibfield  {author} {\bibinfo {author} {\bibfnamefont {T.~D.}\ \bibnamefont
  {K{\"u}hner}}\ and\ \bibinfo {author} {\bibfnamefont {H.}~\bibnamefont
  {Monien}},\ }\bibfield  {title} {\bibinfo {title} {Phases of the
  one-dimensional bose-hubbard model},\ }\href
  {https://doi.org/10.1103/PhysRevB.58.R14741} {\bibfield  {journal} {\bibinfo
  {journal} {Phys. Rev. B}\ }\textbf {\bibinfo {volume} {58}},\ \bibinfo
  {pages} {R14741} (\bibinfo {year} {1998})}\BibitemShut {NoStop}%
\bibitem [{\citenamefont {Greiner}\ \emph {et~al.}(2002)\citenamefont
  {Greiner}, \citenamefont {Mandel}, \citenamefont {Esslinger}, \citenamefont
  {H{\"a}nsch},\ and\ \citenamefont {Bloch}}]{Greiner2002}%
  \BibitemOpen
  \bibfield  {author} {\bibinfo {author} {\bibfnamefont {M.}~\bibnamefont
  {Greiner}}, \bibinfo {author} {\bibfnamefont {O.}~\bibnamefont {Mandel}},
  \bibinfo {author} {\bibfnamefont {T.}~\bibnamefont {Esslinger}}, \bibinfo
  {author} {\bibfnamefont {T.~W.}\ \bibnamefont {H{\"a}nsch}},\ and\ \bibinfo
  {author} {\bibfnamefont {I.}~\bibnamefont {Bloch}},\ }\bibfield  {title}
  {\bibinfo {title} {Quantum phase transition from a superfluid to a mott
  insulator in a gas of ultracold atoms},\ }\href
  {https://doi.org/10.1038/415039a} {\bibfield  {journal} {\bibinfo  {journal}
  {Nature}\ }\textbf {\bibinfo {volume} {415}},\ \bibinfo {pages} {39}
  (\bibinfo {year} {2002})}\BibitemShut {NoStop}%
\bibitem [{\citenamefont {Wang}\ and\ \citenamefont {Hioe}(1973)}]{Wang1973}%
  \BibitemOpen
  \bibfield  {author} {\bibinfo {author} {\bibfnamefont {Y.~K.}\ \bibnamefont
  {Wang}}\ and\ \bibinfo {author} {\bibfnamefont {F.~T.}\ \bibnamefont
  {Hioe}},\ }\bibfield  {title} {\bibinfo {title} {Phase transition in the
  dicke model of superradiance},\ }\href
  {https://doi.org/10.1103/PhysRevA.7.831} {\bibfield  {journal} {\bibinfo
  {journal} {Phys. Rev. A}\ }\textbf {\bibinfo {volume} {7}},\ \bibinfo {pages}
  {831} (\bibinfo {year} {1973})}\BibitemShut {NoStop}%
\bibitem [{\citenamefont {Hwang}\ \emph {et~al.}(2015)\citenamefont {Hwang},
  \citenamefont {Puebla},\ and\ \citenamefont {Plenio}}]{Hwang2015}%
  \BibitemOpen
  \bibfield  {author} {\bibinfo {author} {\bibfnamefont {M.-J.}\ \bibnamefont
  {Hwang}}, \bibinfo {author} {\bibfnamefont {R.}~\bibnamefont {Puebla}},\ and\
  \bibinfo {author} {\bibfnamefont {M.~B.}\ \bibnamefont {Plenio}},\ }\bibfield
   {title} {\bibinfo {title} {Quantum phase transition and universal dynamics
  in the rabi model},\ }\href {https://doi.org/10.1103/PhysRevLett.115.180404}
  {\bibfield  {journal} {\bibinfo  {journal} {Phys. Rev. Lett.}\ }\textbf
  {\bibinfo {volume} {115}},\ \bibinfo {pages} {180404} (\bibinfo {year}
  {2015})}\BibitemShut {NoStop}%
\bibitem [{\citenamefont {Giamarchi}(2003)}]{Giamarchi2003}%
  \BibitemOpen
  \bibfield  {author} {\bibinfo {author} {\bibfnamefont {T.}~\bibnamefont
  {Giamarchi}},\ }\href@noop {} {\emph {\bibinfo {title} {Quantum Physics in
  One Dimension}}}\ (\bibinfo  {publisher} {Oxford University Press},\ \bibinfo
  {address} {Oxford},\ \bibinfo {year} {2003})\BibitemShut {NoStop}%
\bibitem [{\citenamefont {Gogolin}\ \emph {et~al.}(1998)\citenamefont
  {Gogolin}, \citenamefont {Nersesyan},\ and\ \citenamefont
  {Tsvelik}}]{Gogolin1998}%
  \BibitemOpen
  \bibfield  {author} {\bibinfo {author} {\bibfnamefont {A.~O.}\ \bibnamefont
  {Gogolin}}, \bibinfo {author} {\bibfnamefont {A.~A.}\ \bibnamefont
  {Nersesyan}},\ and\ \bibinfo {author} {\bibfnamefont {A.~M.}\ \bibnamefont
  {Tsvelik}},\ }\href@noop {} {\emph {\bibinfo {title} {Bosonization and
  Strongly Correlated Systems}}}\ (\bibinfo  {publisher} {Cambridge University
  Press},\ \bibinfo {address} {New York},\ \bibinfo {year} {1998})\BibitemShut
  {NoStop}%
\bibitem [{\citenamefont {Hofferberth}\ \emph {et~al.}(2007)\citenamefont
  {Hofferberth}, \citenamefont {Lesanovsky}, \citenamefont {Fischer},
  \citenamefont {Schumm},\ and\ \citenamefont
  {Schmiedmayer}}]{Hofferberth2007}%
  \BibitemOpen
  \bibfield  {author} {\bibinfo {author} {\bibfnamefont {S.}~\bibnamefont
  {Hofferberth}}, \bibinfo {author} {\bibfnamefont {I.}~\bibnamefont
  {Lesanovsky}}, \bibinfo {author} {\bibfnamefont {B.}~\bibnamefont {Fischer}},
  \bibinfo {author} {\bibfnamefont {T.}~\bibnamefont {Schumm}},\ and\ \bibinfo
  {author} {\bibfnamefont {J.}~\bibnamefont {Schmiedmayer}},\ }\bibfield
  {title} {\bibinfo {title} {Non-equilibrium coherence dynamics in
  one-dimensional bose gases},\ }\href {https://doi.org/10.1038/nature06149}
  {\bibfield  {journal} {\bibinfo  {journal} {Nature}\ }\textbf {\bibinfo
  {volume} {449}},\ \bibinfo {pages} {324} (\bibinfo {year}
  {2007})}\BibitemShut {NoStop}%
\bibitem [{\citenamefont {Erne}\ \emph {et~al.}(2018)\citenamefont {Erne},
  \citenamefont {B{\"u}cker}, \citenamefont {Gasenzer}, \citenamefont
  {Berges},\ and\ \citenamefont {Schmiedmayer}}]{Erne2018}%
  \BibitemOpen
  \bibfield  {author} {\bibinfo {author} {\bibfnamefont {S.}~\bibnamefont
  {Erne}}, \bibinfo {author} {\bibfnamefont {R.}~\bibnamefont {B{\"u}cker}},
  \bibinfo {author} {\bibfnamefont {T.}~\bibnamefont {Gasenzer}}, \bibinfo
  {author} {\bibfnamefont {J.}~\bibnamefont {Berges}},\ and\ \bibinfo {author}
  {\bibfnamefont {J.}~\bibnamefont {Schmiedmayer}},\ }\bibfield  {title}
  {\bibinfo {title} {Universal dynamics in an isolated one-dimensional bose gas
  far from equilibrium},\ }\href {https://doi.org/10.1038/s41586-018-0667-0}
  {\bibfield  {journal} {\bibinfo  {journal} {Nature}\ }\textbf {\bibinfo
  {volume} {563}},\ \bibinfo {pages} {225} (\bibinfo {year}
  {2018})}\BibitemShut {NoStop}%
\bibitem [{\citenamefont {Haller}\ \emph {et~al.}(2010)\citenamefont {Haller},
  \citenamefont {Hart}, \citenamefont {Mark}, \citenamefont {Danzl},
  \citenamefont {Reichs{\"o}llner}, \citenamefont {Gustavsson}, \citenamefont
  {Dalmonte}, \citenamefont {Pupillo},\ and\ \citenamefont
  {N{\"a}gerl}}]{Haller2010}%
  \BibitemOpen
  \bibfield  {author} {\bibinfo {author} {\bibfnamefont {E.}~\bibnamefont
  {Haller}}, \bibinfo {author} {\bibfnamefont {R.}~\bibnamefont {Hart}},
  \bibinfo {author} {\bibfnamefont {M.~J.}\ \bibnamefont {Mark}}, \bibinfo
  {author} {\bibfnamefont {J.~G.}\ \bibnamefont {Danzl}}, \bibinfo {author}
  {\bibfnamefont {L.}~\bibnamefont {Reichs{\"o}llner}}, \bibinfo {author}
  {\bibfnamefont {M.}~\bibnamefont {Gustavsson}}, \bibinfo {author}
  {\bibfnamefont {M.}~\bibnamefont {Dalmonte}}, \bibinfo {author}
  {\bibfnamefont {G.}~\bibnamefont {Pupillo}},\ and\ \bibinfo {author}
  {\bibfnamefont {H.-C.}\ \bibnamefont {N{\"a}gerl}},\ }\bibfield  {title}
  {\bibinfo {title} {Pinning quantum phase transition for a luttinger liquid of
  strongly interacting bosons},\ }\href {https://doi.org/10.1038/nature09259}
  {\bibfield  {journal} {\bibinfo  {journal} {Nature}\ }\textbf {\bibinfo
  {volume} {466}},\ \bibinfo {pages} {597} (\bibinfo {year}
  {2010})}\BibitemShut {NoStop}%
\bibitem [{\citenamefont {Yang}\ \emph {et~al.}(2017)\citenamefont {Yang},
  \citenamefont {Chen}, \citenamefont {Zheng}, \citenamefont {Sun},
  \citenamefont {Dai}, \citenamefont {Guan}, \citenamefont {Yuan},\ and\
  \citenamefont {Pan}}]{Yang2017}%
  \BibitemOpen
  \bibfield  {author} {\bibinfo {author} {\bibfnamefont {B.}~\bibnamefont
  {Yang}}, \bibinfo {author} {\bibfnamefont {Y.-Y.}\ \bibnamefont {Chen}},
  \bibinfo {author} {\bibfnamefont {Y.-G.}\ \bibnamefont {Zheng}}, \bibinfo
  {author} {\bibfnamefont {H.}~\bibnamefont {Sun}}, \bibinfo {author}
  {\bibfnamefont {H.-N.}\ \bibnamefont {Dai}}, \bibinfo {author} {\bibfnamefont
  {X.-W.}\ \bibnamefont {Guan}}, \bibinfo {author} {\bibfnamefont {Z.-S.}\
  \bibnamefont {Yuan}},\ and\ \bibinfo {author} {\bibfnamefont {J.-W.}\
  \bibnamefont {Pan}},\ }\bibfield  {title} {\bibinfo {title} {Quantum
  criticality and the tomonaga-luttinger liquid in one-dimensional bose
  gases},\ }\href {https://doi.org/10.1103/PhysRevLett.119.165701} {\bibfield
  {journal} {\bibinfo  {journal} {Phys. Rev. Lett.}\ }\textbf {\bibinfo
  {volume} {119}},\ \bibinfo {pages} {165701} (\bibinfo {year}
  {2017})}\BibitemShut {NoStop}%
\bibitem [{\citenamefont {Polkovnikov}\ \emph {et~al.}(2011)\citenamefont
  {Polkovnikov}, \citenamefont {Sengupta}, \citenamefont {Silva},\ and\
  \citenamefont {Vengalattore}}]{Polkovnikov2011}%
  \BibitemOpen
  \bibfield  {author} {\bibinfo {author} {\bibfnamefont {A.}~\bibnamefont
  {Polkovnikov}}, \bibinfo {author} {\bibfnamefont {K.}~\bibnamefont
  {Sengupta}}, \bibinfo {author} {\bibfnamefont {A.}~\bibnamefont {Silva}},\
  and\ \bibinfo {author} {\bibfnamefont {M.}~\bibnamefont {Vengalattore}},\
  }\bibfield  {title} {\bibinfo {title} {Colloquium: Nonequilibrium dynamics of
  closed interacting quantum systems},\ }\href
  {https://doi.org/10.1103/RevModPhys.83.863} {\bibfield  {journal} {\bibinfo
  {journal} {Rev. Mod. Phys.}\ }\textbf {\bibinfo {volume} {83}},\ \bibinfo
  {pages} {863} (\bibinfo {year} {2011})}\BibitemShut {NoStop}%
\bibitem [{\citenamefont {Zvyagin}(2016)}]{Zvyagin2016}%
  \BibitemOpen
  \bibfield  {author} {\bibinfo {author} {\bibfnamefont {A.~A.}\ \bibnamefont
  {Zvyagin}},\ }\bibfield  {title} {\bibinfo {title} {Dynamical quantum phase
  transitions (review article)},\ }\href {https://doi.org/10.1063/1.4969869}
  {\bibfield  {journal} {\bibinfo  {journal} {Low Temp. Phys.}\ }\textbf
  {\bibinfo {volume} {42}},\ \bibinfo {pages} {971} (\bibinfo {year}
  {2016})}\BibitemShut {NoStop}%
\bibitem [{\citenamefont {Heyl}(2018)}]{Heyl2018}%
  \BibitemOpen
  \bibfield  {author} {\bibinfo {author} {\bibfnamefont {M.}~\bibnamefont
  {Heyl}},\ }\bibfield  {title} {\bibinfo {title} {Dynamical quantum phase
  transitions: a review},\ }\href {https://doi.org/10.1088/1361-6633/aaaf9a}
  {\bibfield  {journal} {\bibinfo  {journal} {Rep. Prog. Phys.}\ }\textbf
  {\bibinfo {volume} {81}},\ \bibinfo {pages} {054001} (\bibinfo {year}
  {2018})}\BibitemShut {NoStop}%
\bibitem [{\citenamefont {Sengupta}\ \emph {et~al.}(2004)\citenamefont
  {Sengupta}, \citenamefont {Powell},\ and\ \citenamefont
  {Sachdev}}]{Sengupta2004}%
  \BibitemOpen
  \bibfield  {author} {\bibinfo {author} {\bibfnamefont {K.}~\bibnamefont
  {Sengupta}}, \bibinfo {author} {\bibfnamefont {S.}~\bibnamefont {Powell}},\
  and\ \bibinfo {author} {\bibfnamefont {S.}~\bibnamefont {Sachdev}},\
  }\bibfield  {title} {\bibinfo {title} {Quench dynamics across quantum
  critical points},\ }\href {https://doi.org/10.1103/PhysRevA.69.053616}
  {\bibfield  {journal} {\bibinfo  {journal} {Phys. Rev. A}\ }\textbf {\bibinfo
  {volume} {69}},\ \bibinfo {pages} {053616} (\bibinfo {year}
  {2004})}\BibitemShut {NoStop}%
\bibitem [{\citenamefont {Zurek}\ \emph {et~al.}(2005)\citenamefont {Zurek},
  \citenamefont {Dorner},\ and\ \citenamefont {Zoller}}]{Zurek2005}%
  \BibitemOpen
  \bibfield  {author} {\bibinfo {author} {\bibfnamefont {W.~H.}\ \bibnamefont
  {Zurek}}, \bibinfo {author} {\bibfnamefont {U.}~\bibnamefont {Dorner}},\ and\
  \bibinfo {author} {\bibfnamefont {P.}~\bibnamefont {Zoller}},\ }\bibfield
  {title} {\bibinfo {title} {Dynamics of a quantum phase transition},\ }\href
  {https://doi.org/10.1103/PhysRevLett.95.105701} {\bibfield  {journal}
  {\bibinfo  {journal} {Phys. Rev. Lett.}\ }\textbf {\bibinfo {volume} {95}},\
  \bibinfo {pages} {105701} (\bibinfo {year} {2005})}\BibitemShut {NoStop}%
\bibitem [{\citenamefont {Kollath}\ \emph {et~al.}(2007)\citenamefont
  {Kollath}, \citenamefont {L{\"a}uchli},\ and\ \citenamefont
  {Altman}}]{Kollath2007}%
  \BibitemOpen
  \bibfield  {author} {\bibinfo {author} {\bibfnamefont {C.}~\bibnamefont
  {Kollath}}, \bibinfo {author} {\bibfnamefont {A.~M.}\ \bibnamefont
  {L{\"a}uchli}},\ and\ \bibinfo {author} {\bibfnamefont {E.}~\bibnamefont
  {Altman}},\ }\bibfield  {title} {\bibinfo {title} {Quench dynamics and
  nonequilibrium phase diagram of the bose-hubbard model},\ }\href
  {https://doi.org/10.1103/PhysRevLett.98.180601} {\bibfield  {journal}
  {\bibinfo  {journal} {Phys. Rev. Lett.}\ }\textbf {\bibinfo {volume} {98}},\
  \bibinfo {pages} {180601} (\bibinfo {year} {2007})}\BibitemShut {NoStop}%
\bibitem [{\citenamefont {Braun}\ \emph {et~al.}(2015)\citenamefont {Braun},
  \citenamefont {Friesdorf}, \citenamefont {Hodgman}, \citenamefont
  {Schreiber}, \citenamefont {Ronzheimer}, \citenamefont {Riera}, \citenamefont
  {del Rey}, \citenamefont {Bloch}, \citenamefont {Eisert},\ and\ \citenamefont
  {Schneider}}]{Braun2015}%
  \BibitemOpen
  \bibfield  {author} {\bibinfo {author} {\bibfnamefont {S.}~\bibnamefont
  {Braun}}, \bibinfo {author} {\bibfnamefont {M.}~\bibnamefont {Friesdorf}},
  \bibinfo {author} {\bibfnamefont {S.~S.}\ \bibnamefont {Hodgman}}, \bibinfo
  {author} {\bibfnamefont {M.}~\bibnamefont {Schreiber}}, \bibinfo {author}
  {\bibfnamefont {J.~P.}\ \bibnamefont {Ronzheimer}}, \bibinfo {author}
  {\bibfnamefont {A.}~\bibnamefont {Riera}}, \bibinfo {author} {\bibfnamefont
  {M.}~\bibnamefont {del Rey}}, \bibinfo {author} {\bibfnamefont
  {I.}~\bibnamefont {Bloch}}, \bibinfo {author} {\bibfnamefont
  {J.}~\bibnamefont {Eisert}},\ and\ \bibinfo {author} {\bibfnamefont
  {U.}~\bibnamefont {Schneider}},\ }\bibfield  {title} {\bibinfo {title}
  {Emergence of coherence and the dynamics of quantum phase transitions},\
  }\href {https://doi.org/10.1073/pnas.1408861112} {\bibfield  {journal}
  {\bibinfo  {journal} {Proc. Natl. Acad. Sci.}\ }\textbf {\bibinfo {volume}
  {112}},\ \bibinfo {pages} {3641} (\bibinfo {year} {2015})}\BibitemShut
  {NoStop}%
\bibitem [{\citenamefont {Sch{\"u}tzhold}\ \emph {et~al.}(2006)\citenamefont
  {Sch{\"u}tzhold}, \citenamefont {Uhlmann}, \citenamefont {Xu},\ and\
  \citenamefont {Fischer}}]{Schuetzhold2006}%
  \BibitemOpen
  \bibfield  {author} {\bibinfo {author} {\bibfnamefont {R.}~\bibnamefont
  {Sch{\"u}tzhold}}, \bibinfo {author} {\bibfnamefont {M.}~\bibnamefont
  {Uhlmann}}, \bibinfo {author} {\bibfnamefont {Y.}~\bibnamefont {Xu}},\ and\
  \bibinfo {author} {\bibfnamefont {U.~R.}\ \bibnamefont {Fischer}},\
  }\bibfield  {title} {\bibinfo {title} {Sweeping from the superfluid to the
  mott phase in the bose-hubbard model},\ }\href
  {https://doi.org/10.1103/PhysRevLett.97.200601} {\bibfield  {journal}
  {\bibinfo  {journal} {Phys. Rev. Lett.}\ }\textbf {\bibinfo {volume} {97}},\
  \bibinfo {pages} {200601} (\bibinfo {year} {2006})}\BibitemShut {NoStop}%
\bibitem [{\citenamefont {Diehl}\ \emph {et~al.}(2010)\citenamefont {Diehl},
  \citenamefont {Tomadin}, \citenamefont {Micheli}, \citenamefont {Fazio},\
  and\ \citenamefont {Zoller}}]{Diehl2010}%
  \BibitemOpen
  \bibfield  {author} {\bibinfo {author} {\bibfnamefont {S.}~\bibnamefont
  {Diehl}}, \bibinfo {author} {\bibfnamefont {A.}~\bibnamefont {Tomadin}},
  \bibinfo {author} {\bibfnamefont {A.}~\bibnamefont {Micheli}}, \bibinfo
  {author} {\bibfnamefont {R.}~\bibnamefont {Fazio}},\ and\ \bibinfo {author}
  {\bibfnamefont {P.}~\bibnamefont {Zoller}},\ }\bibfield  {title} {\bibinfo
  {title} {Dynamical phase transitions and instabilities in open atomic
  many-body systems},\ }\href {https://doi.org/10.1103/physrevlett.105.015702}
  {\bibfield  {journal} {\bibinfo  {journal} {Phys. Rev. Lett.}\ }\textbf
  {\bibinfo {volume} {105}},\ \bibinfo {pages} {015702} (\bibinfo {year}
  {2010})}\BibitemShut {NoStop}%
\bibitem [{\citenamefont {Cazalilla}\ \emph {et~al.}(2011)\citenamefont
  {Cazalilla}, \citenamefont {Citro}, \citenamefont {Giamarchi}, \citenamefont
  {Orignac},\ and\ \citenamefont {Rigol}}]{Cazalilla2011}%
  \BibitemOpen
  \bibfield  {author} {\bibinfo {author} {\bibfnamefont {M.~A.}\ \bibnamefont
  {Cazalilla}}, \bibinfo {author} {\bibfnamefont {R.}~\bibnamefont {Citro}},
  \bibinfo {author} {\bibfnamefont {T.}~\bibnamefont {Giamarchi}}, \bibinfo
  {author} {\bibfnamefont {E.}~\bibnamefont {Orignac}},\ and\ \bibinfo {author}
  {\bibfnamefont {M.}~\bibnamefont {Rigol}},\ }\bibfield  {title} {\bibinfo
  {title} {One dimensional bosons: From condensed matter systems to ultracold
  gases},\ }\href {https://doi.org/10.1103/RevModPhys.83.1405} {\bibfield
  {journal} {\bibinfo  {journal} {Rev. Mod. Phys.}\ }\textbf {\bibinfo {volume}
  {83}},\ \bibinfo {pages} {1405} (\bibinfo {year} {2011})}\BibitemShut
  {NoStop}%
\bibitem [{\citenamefont {Heyl}\ \emph {et~al.}(2013)\citenamefont {Heyl},
  \citenamefont {Polkovnikov},\ and\ \citenamefont {Kehrein}}]{Heyl2013}%
  \BibitemOpen
  \bibfield  {author} {\bibinfo {author} {\bibfnamefont {M.}~\bibnamefont
  {Heyl}}, \bibinfo {author} {\bibfnamefont {A.}~\bibnamefont {Polkovnikov}},\
  and\ \bibinfo {author} {\bibfnamefont {S.}~\bibnamefont {Kehrein}},\
  }\bibfield  {title} {\bibinfo {title} {Dynamical quantum phase transitions in
  the transverse-field ising model},\ }\href
  {https://doi.org/10.1103/physrevlett.110.135704} {\bibfield  {journal}
  {\bibinfo  {journal} {Phys. Rev. Lett.}\ }\textbf {\bibinfo {volume} {110}},\
  \bibinfo {pages} {135704} (\bibinfo {year} {2013})}\BibitemShut {NoStop}%
\bibitem [{\citenamefont {Huang}\ and\ \citenamefont
  {Balatsky}(2016)}]{Huang2016}%
  \BibitemOpen
  \bibfield  {author} {\bibinfo {author} {\bibfnamefont {Z.}~\bibnamefont
  {Huang}}\ and\ \bibinfo {author} {\bibfnamefont {A.~V.}\ \bibnamefont
  {Balatsky}},\ }\bibfield  {title} {\bibinfo {title} {Dynamical quantum phase
  transitions: Role of topological nodes in wave function overlaps},\ }\href
  {https://doi.org/10.1103/PhysRevLett.117.086802} {\bibfield  {journal}
  {\bibinfo  {journal} {Phys. Rev. Lett.}\ }\textbf {\bibinfo {volume} {117}},\
  \bibinfo {pages} {086802} (\bibinfo {year} {2016})}\BibitemShut {NoStop}%
\bibitem [{\citenamefont {Halimeh}\ and\ \citenamefont
  {Zauner-Stauber}(2017)}]{Halimeh2017}%
  \BibitemOpen
  \bibfield  {author} {\bibinfo {author} {\bibfnamefont {J.~C.}\ \bibnamefont
  {Halimeh}}\ and\ \bibinfo {author} {\bibfnamefont {V.}~\bibnamefont
  {Zauner-Stauber}},\ }\bibfield  {title} {\bibinfo {title} {Dynamical phase
  diagram of quantum spin chains with long-range interactions},\ }\href
  {https://doi.org/10.1103/physrevb.96.134427} {\bibfield  {journal} {\bibinfo
  {journal} {Phys. Rev. B}\ }\textbf {\bibinfo {volume} {96}},\ \bibinfo
  {pages} {134427} (\bibinfo {year} {2017})}\BibitemShut {NoStop}%
\bibitem [{\citenamefont {Jurcevic}\ \emph {et~al.}(2017)\citenamefont
  {Jurcevic}, \citenamefont {Shen}, \citenamefont {Hauke}, \citenamefont
  {Maier}, \citenamefont {Brydges}, \citenamefont {Hempel}, \citenamefont
  {Lanyon}, \citenamefont {Heyl}, \citenamefont {Blatt},\ and\ \citenamefont
  {Roos}}]{Jurcevic2017}%
  \BibitemOpen
  \bibfield  {author} {\bibinfo {author} {\bibfnamefont {P.}~\bibnamefont
  {Jurcevic}}, \bibinfo {author} {\bibfnamefont {H.}~\bibnamefont {Shen}},
  \bibinfo {author} {\bibfnamefont {P.}~\bibnamefont {Hauke}}, \bibinfo
  {author} {\bibfnamefont {C.}~\bibnamefont {Maier}}, \bibinfo {author}
  {\bibfnamefont {T.}~\bibnamefont {Brydges}}, \bibinfo {author} {\bibfnamefont
  {C.}~\bibnamefont {Hempel}}, \bibinfo {author} {\bibfnamefont {B.~P.}\
  \bibnamefont {Lanyon}}, \bibinfo {author} {\bibfnamefont {M.}~\bibnamefont
  {Heyl}}, \bibinfo {author} {\bibfnamefont {R.}~\bibnamefont {Blatt}},\ and\
  \bibinfo {author} {\bibfnamefont {C.~F.}\ \bibnamefont {Roos}},\ }\bibfield
  {title} {\bibinfo {title} {Direct observation of dynamical quantum phase
  transitions in an interacting many-body system},\ }\href
  {https://doi.org/10.1103/physrevlett.119.080501} {\bibfield  {journal}
  {\bibinfo  {journal} {Phys. Rev. Lett.}\ }\textbf {\bibinfo {volume} {119}},\
  \bibinfo {pages} {080501} (\bibinfo {year} {2017})}\BibitemShut {NoStop}%
\bibitem [{\citenamefont {{\v{Z}}unkovi{\v{c}}}\ \emph
  {et~al.}(2018)\citenamefont {{\v{Z}}unkovi{\v{c}}}, \citenamefont {Heyl},
  \citenamefont {Knap},\ and\ \citenamefont {Silva}}]{Zunkovic2018}%
  \BibitemOpen
  \bibfield  {author} {\bibinfo {author} {\bibfnamefont {B.}~\bibnamefont
  {{\v{Z}}unkovi{\v{c}}}}, \bibinfo {author} {\bibfnamefont {M.}~\bibnamefont
  {Heyl}}, \bibinfo {author} {\bibfnamefont {M.}~\bibnamefont {Knap}},\ and\
  \bibinfo {author} {\bibfnamefont {A.}~\bibnamefont {Silva}},\ }\bibfield
  {title} {\bibinfo {title} {Dynamical quantum phase transitions in spin chains
  with long-range interactions: Merging different concepts of nonequilibrium
  criticality},\ }\href {https://doi.org/10.1103/physrevlett.120.130601}
  {\bibfield  {journal} {\bibinfo  {journal} {Phys. Rev. Lett.}\ }\textbf
  {\bibinfo {volume} {120}},\ \bibinfo {pages} {130601} (\bibinfo {year}
  {2018})}\BibitemShut {NoStop}%
\bibitem [{\citenamefont {Jafari}(2019)}]{Jafari2019}%
  \BibitemOpen
  \bibfield  {author} {\bibinfo {author} {\bibfnamefont {R.}~\bibnamefont
  {Jafari}},\ }\bibfield  {title} {\bibinfo {title} {Dynamical quantum phase
  transition and quasi particle excitation},\ }\href
  {https://doi.org/10.1038/s41598-019-39595-3} {\bibfield  {journal} {\bibinfo
  {journal} {Sci. Rep.}\ }\textbf {\bibinfo {volume} {9}},\ \bibinfo {pages}
  {2871} (\bibinfo {year} {2019})}\BibitemShut {NoStop}%
\bibitem [{\citenamefont {Lacki}\ and\ \citenamefont {Heyl}(2019)}]{Lacki2019}%
  \BibitemOpen
  \bibfield  {author} {\bibinfo {author} {\bibfnamefont {M.}~\bibnamefont
  {Lacki}}\ and\ \bibinfo {author} {\bibfnamefont {M.}~\bibnamefont {Heyl}},\
  }\bibfield  {title} {\bibinfo {title} {Dynamical quantum phase transitions in
  collapse and revival oscillations of a quenched superfluid},\ }\href
  {https://doi.org/10.1103/PhysRevB.99.121107} {\bibfield  {journal} {\bibinfo
  {journal} {Phys. Rev. B}\ }\textbf {\bibinfo {volume} {99}},\ \bibinfo
  {pages} {121107(R)} (\bibinfo {year} {2019})}\BibitemShut {NoStop}%
\bibitem [{\citenamefont {Hickey}\ \emph {et~al.}(2014)\citenamefont {Hickey},
  \citenamefont {Genway},\ and\ \citenamefont {Garrahan}}]{Hickey2014}%
  \BibitemOpen
  \bibfield  {author} {\bibinfo {author} {\bibfnamefont {J.~M.}\ \bibnamefont
  {Hickey}}, \bibinfo {author} {\bibfnamefont {S.}~\bibnamefont {Genway}},\
  and\ \bibinfo {author} {\bibfnamefont {J.~P.}\ \bibnamefont {Garrahan}},\
  }\bibfield  {title} {\bibinfo {title} {Dynamical phase transitions,
  time-integrated observables, and geometry of states},\ }\href
  {https://doi.org/10.1103/PhysRevB.89.054301} {\bibfield  {journal} {\bibinfo
  {journal} {Phys. Rev. B}\ }\textbf {\bibinfo {volume} {89}},\ \bibinfo
  {pages} {054301} (\bibinfo {year} {2014})}\BibitemShut {NoStop}%
\bibitem [{\citenamefont {Zauner-Stauber}\ and\ \citenamefont
  {Halimeh}(2017)}]{Zauner2017}%
  \BibitemOpen
  \bibfield  {author} {\bibinfo {author} {\bibfnamefont {V.}~\bibnamefont
  {Zauner-Stauber}}\ and\ \bibinfo {author} {\bibfnamefont {J.~C.}\
  \bibnamefont {Halimeh}},\ }\bibfield  {title} {\bibinfo {title} {Probing the
  anomalous dynamical phase in long-range quantum spin chains through
  fisher-zero lines},\ }\href {https://doi.org/10.1103/PhysRevE.96.062118}
  {\bibfield  {journal} {\bibinfo  {journal} {Phys. Rev. E}\ }\textbf {\bibinfo
  {volume} {96}},\ \bibinfo {pages} {062118} (\bibinfo {year}
  {2017})}\BibitemShut {NoStop}%
\bibitem [{\citenamefont {Recati}\ \emph {et~al.}(2005)\citenamefont {Recati},
  \citenamefont {Fedichev}, \citenamefont {Zwerger}, \citenamefont {von
  Delft},\ and\ \citenamefont {Zoller}}]{Recati2005}%
  \BibitemOpen
  \bibfield  {author} {\bibinfo {author} {\bibfnamefont {A.}~\bibnamefont
  {Recati}}, \bibinfo {author} {\bibfnamefont {P.~O.}\ \bibnamefont
  {Fedichev}}, \bibinfo {author} {\bibfnamefont {W.}~\bibnamefont {Zwerger}},
  \bibinfo {author} {\bibfnamefont {J.}~\bibnamefont {von Delft}},\ and\
  \bibinfo {author} {\bibfnamefont {P.}~\bibnamefont {Zoller}},\ }\bibfield
  {title} {\bibinfo {title} {Atomic quantum dots coupled to a reservoir of a
  superfluid bose-einstein condensate},\ }\href
  {https://doi.org/10.1103/PhysRevLett.94.040404} {\bibfield  {journal}
  {\bibinfo  {journal} {Phys. Rev. Lett.}\ }\textbf {\bibinfo {volume} {94}},\
  \bibinfo {pages} {040404} (\bibinfo {year} {2005})}\BibitemShut {NoStop}%
\bibitem [{\citenamefont {Do}\ \emph {et~al.}(2019)\citenamefont {Do},
  \citenamefont {Gessner}, \citenamefont {Cataliotti},\ and\ \citenamefont
  {Smerzi}}]{Do2019}%
  \BibitemOpen
  \bibfield  {author} {\bibinfo {author} {\bibfnamefont {H.~V.}\ \bibnamefont
  {Do}}, \bibinfo {author} {\bibfnamefont {M.}~\bibnamefont {Gessner}},
  \bibinfo {author} {\bibfnamefont {F.~S.}\ \bibnamefont {Cataliotti}},\ and\
  \bibinfo {author} {\bibfnamefont {A.}~\bibnamefont {Smerzi}},\ }\bibfield
  {title} {\bibinfo {title} {Measuring geometric phases with a dynamical
  quantum zeno effect in a bose-einstein condensate},\ }\href
  {https://arxiv.org/abs/1903.05122} {\bibfield  {journal} {\bibinfo  {journal}
  {arXiv:1903.05122 [quant-ph]}\ } (\bibinfo {year} {2019})}\BibitemShut
  {NoStop}%
\bibitem [{\citenamefont {Fedichev}\ and\ \citenamefont
  {Fischer}(2003)}]{Fedichev2003}%
  \BibitemOpen
  \bibfield  {author} {\bibinfo {author} {\bibfnamefont {P.~O.}\ \bibnamefont
  {Fedichev}}\ and\ \bibinfo {author} {\bibfnamefont {U.~R.}\ \bibnamefont
  {Fischer}},\ }\bibfield  {title} {\bibinfo {title} {Gibbons-hawking effect in
  the sonic de sitter space-time of an expanding bose-einstein-condensed gas},\
  }\href {https://doi.org/10.1103/PhysRevLett.91.240407} {\bibfield  {journal}
  {\bibinfo  {journal} {Phys. Rev. Lett.}\ }\textbf {\bibinfo {volume} {91}},\
  \bibinfo {pages} {240407} (\bibinfo {year} {2003})}\BibitemShut {NoStop}%
\bibitem [{\citenamefont {West}(2013)}]{West2013}%
  \BibitemOpen
  \bibfield  {author} {\bibinfo {author} {\bibfnamefont {T.}~\bibnamefont
  {West}},\ }\emph {\bibinfo {title} {Quantum Dot Dynamics in a Bose-Einstein
  Condensate}},\ \href {http://hdl.handle.net/10044/1/23993} {Ph.D. thesis},\
  \bibinfo  {school} {Imperial College London} (\bibinfo {year}
  {2013})\BibitemShut {NoStop}%
\bibitem [{\citenamefont {Bruderer}\ and\ \citenamefont
  {Jaksch}(2006)}]{Bruderer2006}%
  \BibitemOpen
  \bibfield  {author} {\bibinfo {author} {\bibfnamefont {M.}~\bibnamefont
  {Bruderer}}\ and\ \bibinfo {author} {\bibfnamefont {D.}~\bibnamefont
  {Jaksch}},\ }\bibfield  {title} {\bibinfo {title} {Probing bec phase
  fluctuations with atomic quantum dots},\ }\href
  {https://doi.org/10.1088/1367-2630/8/6/087} {\bibfield  {journal} {\bibinfo
  {journal} {New J. Phys.}\ }\textbf {\bibinfo {volume} {8}},\ \bibinfo {pages}
  {87} (\bibinfo {year} {2006})}\BibitemShut {NoStop}%
\bibitem [{\citenamefont {Abdi}\ and\ \citenamefont
  {Plenio}(2018)}]{Abdi2018a}%
  \BibitemOpen
  \bibfield  {author} {\bibinfo {author} {\bibfnamefont {M.}~\bibnamefont
  {Abdi}}\ and\ \bibinfo {author} {\bibfnamefont {M.~B.}\ \bibnamefont
  {Plenio}},\ }\bibfield  {title} {\bibinfo {title} {Analog quantum simulation
  of extremely sub-ohmic spin-boson models},\ }\href
  {https://doi.org/10.1103/physreva.98.040303} {\bibfield  {journal} {\bibinfo
  {journal} {Phys. Rev. A}\ }\textbf {\bibinfo {volume} {98}},\ \bibinfo
  {pages} {040303(R)} (\bibinfo {year} {2018})}\BibitemShut {NoStop}%
\bibitem [{\citenamefont {Abdi}\ \emph {et~al.}(2018)\citenamefont {Abdi},
  \citenamefont {Chou}, \citenamefont {Gali},\ and\ \citenamefont
  {Plenio}}]{Abdi2018}%
  \BibitemOpen
  \bibfield  {author} {\bibinfo {author} {\bibfnamefont {M.}~\bibnamefont
  {Abdi}}, \bibinfo {author} {\bibfnamefont {J.-P.}\ \bibnamefont {Chou}},
  \bibinfo {author} {\bibfnamefont {A.}~\bibnamefont {Gali}},\ and\ \bibinfo
  {author} {\bibfnamefont {M.~B.}\ \bibnamefont {Plenio}},\ }\bibfield  {title}
  {\bibinfo {title} {Color centers in hexagonal boron nitride monolayers: A
  group theory and ab initio analysis},\ }\href
  {https://doi.org/10.1021/acsphotonics.7b01442} {\bibfield  {journal}
  {\bibinfo  {journal} {{ACS} Photonics}\ }\textbf {\bibinfo {volume} {5}},\
  \bibinfo {pages} {1967} (\bibinfo {year} {2018})}\BibitemShut {NoStop}%
\bibitem [{\citenamefont {Haldane}(1981)}]{Haldane1981}%
  \BibitemOpen
  \bibfield  {author} {\bibinfo {author} {\bibfnamefont {F.~D.~M.}\
  \bibnamefont {Haldane}},\ }\bibfield  {title} {\bibinfo {title} {Effective
  harmonic-fluid approach to low-energy properties of one-dimensional quantum
  fluids},\ }\href {https://doi.org/10.1103/PhysRevLett.47.1840} {\bibfield
  {journal} {\bibinfo  {journal} {Phys. Rev. Lett.}\ }\textbf {\bibinfo
  {volume} {47}},\ \bibinfo {pages} {1840} (\bibinfo {year}
  {1981})}\BibitemShut {NoStop}%
\bibitem [{\citenamefont {Bennett}\ \emph {et~al.}(2012)\citenamefont
  {Bennett}, \citenamefont {Kolkowitz}, \citenamefont {Unterreithmeier},
  \citenamefont {Rabl}, \citenamefont {Jayich}, \citenamefont {Harris},\ and\
  \citenamefont {Lukin}}]{Bennett2012}%
  \BibitemOpen
  \bibfield  {author} {\bibinfo {author} {\bibfnamefont {S.~D.}\ \bibnamefont
  {Bennett}}, \bibinfo {author} {\bibfnamefont {S.}~\bibnamefont {Kolkowitz}},
  \bibinfo {author} {\bibfnamefont {Q.~P.}\ \bibnamefont {Unterreithmeier}},
  \bibinfo {author} {\bibfnamefont {P.}~\bibnamefont {Rabl}}, \bibinfo {author}
  {\bibfnamefont {A.~C.~B.}\ \bibnamefont {Jayich}}, \bibinfo {author}
  {\bibfnamefont {J.~G.~E.}\ \bibnamefont {Harris}},\ and\ \bibinfo {author}
  {\bibfnamefont {M.~D.}\ \bibnamefont {Lukin}},\ }\bibfield  {title} {\bibinfo
  {title} {Measuring mechanical motion with a single spin},\ }\href
  {https://doi.org/10.1088/1367-2630/14/12/125004} {\bibfield  {journal}
  {\bibinfo  {journal} {New J. Phys.}\ }\textbf {\bibinfo {volume} {14}},\
  \bibinfo {pages} {125004} (\bibinfo {year} {2012})}\BibitemShut {NoStop}%
\bibitem [{\citenamefont {Asadian}\ and\ \citenamefont
  {Abdi}(2016)}]{Asadian2016}%
  \BibitemOpen
  \bibfield  {author} {\bibinfo {author} {\bibfnamefont {A.}~\bibnamefont
  {Asadian}}\ and\ \bibinfo {author} {\bibfnamefont {M.}~\bibnamefont {Abdi}},\
  }\bibfield  {title} {\bibinfo {title} {Heralded entangled coherent states
  between spatially separated massive resonators},\ }\href
  {https://doi.org/10.1103/PhysRevA.93.052315} {\bibfield  {journal} {\bibinfo
  {journal} {Phys. Rev. A}\ }\textbf {\bibinfo {volume} {93}},\ \bibinfo
  {pages} {052315} (\bibinfo {year} {2016})}\BibitemShut {NoStop}%
\bibitem [{\citenamefont {Asadian}\ \emph {et~al.}(2014)\citenamefont
  {Asadian}, \citenamefont {Brukner},\ and\ \citenamefont
  {Rabl}}]{Asadian2014}%
  \BibitemOpen
  \bibfield  {author} {\bibinfo {author} {\bibfnamefont {A.}~\bibnamefont
  {Asadian}}, \bibinfo {author} {\bibfnamefont {C.}~\bibnamefont {Brukner}},\
  and\ \bibinfo {author} {\bibfnamefont {P.}~\bibnamefont {Rabl}},\ }\bibfield
  {title} {\bibinfo {title} {Probing macroscopic realism via ramsey correlation
  measurements},\ }\href {https://doi.org/10.1103/physrevlett.112.190402}
  {\bibfield  {journal} {\bibinfo  {journal} {Phys. Rev. Lett.}\ }\textbf
  {\bibinfo {volume} {112}},\ \bibinfo {pages} {190402} (\bibinfo {year}
  {2014})}\BibitemShut {NoStop}%
\bibitem [{\citenamefont {Carollo}\ and\ \citenamefont
  {Pachos}(2005)}]{Carollo2005}%
  \BibitemOpen
  \bibfield  {author} {\bibinfo {author} {\bibfnamefont {A.~C.~M.}\
  \bibnamefont {Carollo}}\ and\ \bibinfo {author} {\bibfnamefont {J.~K.}\
  \bibnamefont {Pachos}},\ }\bibfield  {title} {\bibinfo {title} {Geometric
  phases and criticality in spin-chain systems},\ }\href
  {https://doi.org/10.1103/physrevlett.95.157203} {\bibfield  {journal}
  {\bibinfo  {journal} {Phys. Rev. Lett.}\ }\textbf {\bibinfo {volume} {95}},\
  \bibinfo {pages} {157203} (\bibinfo {year} {2005})}\BibitemShut {NoStop}%
\bibitem [{\citenamefont {Zhu}(2006)}]{Zhu2006}%
  \BibitemOpen
  \bibfield  {author} {\bibinfo {author} {\bibfnamefont {S.-L.}\ \bibnamefont
  {Zhu}},\ }\bibfield  {title} {\bibinfo {title} {Scaling of geometric phases
  close to the quantum phase transition in {theXYSpin} chain},\ }\href
  {https://doi.org/10.1103/physrevlett.96.077206} {\bibfield  {journal}
  {\bibinfo  {journal} {Phys. Rev. Lett.}\ }\textbf {\bibinfo {volume} {96}},\
  \bibinfo {pages} {077206} (\bibinfo {year} {2006})}\BibitemShut {NoStop}%
\bibitem [{\citenamefont {Hamma}(2006)}]{Hamma2006}%
  \BibitemOpen
  \bibfield  {author} {\bibinfo {author} {\bibfnamefont {A.}~\bibnamefont
  {Hamma}},\ }\bibfield  {title} {\bibinfo {title} {Berry phases and quantum
  phase transitions},\ }\href {https://arxiv.org/abs/quant-ph/0602091}
  {\bibfield  {journal} {\bibinfo  {journal} {arXiv:quant-ph/0602091}\ }
  (\bibinfo {year} {2006})}\BibitemShut {NoStop}%
\bibitem [{\citenamefont {Fujikawa}(2005)}]{Fujikawa2005}%
  \BibitemOpen
  \bibfield  {author} {\bibinfo {author} {\bibfnamefont {K.}~\bibnamefont
  {Fujikawa}},\ }\bibfield  {title} {\bibinfo {title} {Topological properties
  of berry's phase},\ }\href {https://doi.org/10.1142/S0217732305016579}
  {\bibfield  {journal} {\bibinfo  {journal} {Mod. Phys. Lett. A}\ }\textbf
  {\bibinfo {volume} {20}},\ \bibinfo {pages} {335} (\bibinfo {year}
  {2005})}\BibitemShut {NoStop}%
\bibitem [{\citenamefont {Yuan}\ \emph {et~al.}(2007)\citenamefont {Yuan},
  \citenamefont {Zhang},\ and\ \citenamefont {Li}}]{Yuan2007}%
  \BibitemOpen
  \bibfield  {author} {\bibinfo {author} {\bibfnamefont {Z.-G.}\ \bibnamefont
  {Yuan}}, \bibinfo {author} {\bibfnamefont {P.}~\bibnamefont {Zhang}},\ and\
  \bibinfo {author} {\bibfnamefont {S.-S.}\ \bibnamefont {Li}},\ }\bibfield
  {title} {\bibinfo {title} {Loschmidt echo and berry phase of a quantum system
  coupled to an xy spin chain: Proximity to a quantum phase transition},\
  }\href {https://doi.org/PhysRevA.75.012102} {\bibfield  {journal} {\bibinfo
  {journal} {Phys. Rev. A}\ }\textbf {\bibinfo {volume} {75}},\ \bibinfo
  {pages} {012102} (\bibinfo {year} {2007})}\BibitemShut {NoStop}%
\bibitem [{\citenamefont {Tran}\ \emph {et~al.}(2015)\citenamefont {Tran},
  \citenamefont {Bray}, \citenamefont {Ford}, \citenamefont {Toth},\ and\
  \citenamefont {Aharonovich}}]{Tran2015}%
  \BibitemOpen
  \bibfield  {author} {\bibinfo {author} {\bibfnamefont {T.~T.}\ \bibnamefont
  {Tran}}, \bibinfo {author} {\bibfnamefont {K.}~\bibnamefont {Bray}}, \bibinfo
  {author} {\bibfnamefont {M.~J.}\ \bibnamefont {Ford}}, \bibinfo {author}
  {\bibfnamefont {M.}~\bibnamefont {Toth}},\ and\ \bibinfo {author}
  {\bibfnamefont {I.}~\bibnamefont {Aharonovich}},\ }\bibfield  {title}
  {\bibinfo {title} {Quantum emission from hexagonal boron nitride
  monolayers},\ }\href {https://doi.org/10.1038/nnano.2015.242} {\bibfield
  {journal} {\bibinfo  {journal} {Nat. Nanotechnol.}\ }\textbf {\bibinfo
  {volume} {11}},\ \bibinfo {pages} {37} (\bibinfo {year} {2015})}\BibitemShut
  {NoStop}%
\bibitem [{\citenamefont {Abdi}\ \emph {et~al.}(2017)\citenamefont {Abdi},
  \citenamefont {Hwang}, \citenamefont {Aghtar},\ and\ \citenamefont
  {Plenio}}]{Abdi2017}%
  \BibitemOpen
  \bibfield  {author} {\bibinfo {author} {\bibfnamefont {M.}~\bibnamefont
  {Abdi}}, \bibinfo {author} {\bibfnamefont {M.-J.}\ \bibnamefont {Hwang}},
  \bibinfo {author} {\bibfnamefont {M.}~\bibnamefont {Aghtar}},\ and\ \bibinfo
  {author} {\bibfnamefont {M.~B.}\ \bibnamefont {Plenio}},\ }\bibfield  {title}
  {\bibinfo {title} {Spin-mechanical scheme with color centers in hexagonal
  boron nitride membranes},\ }\href
  {https://doi.org/10.1103/physrevlett.119.233602} {\bibfield  {journal}
  {\bibinfo  {journal} {Phys. Rev. Lett.}\ }\textbf {\bibinfo {volume} {119}},\
  \bibinfo {pages} {233602} (\bibinfo {year} {2017})}\BibitemShut {NoStop}%
\bibitem [{\citenamefont {Landau}\ and\ \citenamefont
  {Lifshitz}(1975)}]{Landau1975}%
  \BibitemOpen
  \bibfield  {author} {\bibinfo {author} {\bibfnamefont {L.~D.}\ \bibnamefont
  {Landau}}\ and\ \bibinfo {author} {\bibfnamefont {E.~M.}\ \bibnamefont
  {Lifshitz}},\ }\href@noop {} {\emph {\bibinfo {title} {Theory of
  Elasticity}}}\ (\bibinfo  {publisher} {Pergamon Press},\ \bibinfo {year}
  {1975})\BibitemShut {NoStop}%
\bibitem [{\citenamefont {Abdi}\ \emph {et~al.}(2016)\citenamefont {Abdi},
  \citenamefont {Degenfeld-Schonburg}, \citenamefont {Sameti}, \citenamefont
  {Navarrete-Benlloch},\ and\ \citenamefont {Hartmann}}]{Abdi2016}%
  \BibitemOpen
  \bibfield  {author} {\bibinfo {author} {\bibfnamefont {M.}~\bibnamefont
  {Abdi}}, \bibinfo {author} {\bibfnamefont {P.}~\bibnamefont
  {Degenfeld-Schonburg}}, \bibinfo {author} {\bibfnamefont {M.}~\bibnamefont
  {Sameti}}, \bibinfo {author} {\bibfnamefont {C.}~\bibnamefont
  {Navarrete-Benlloch}},\ and\ \bibinfo {author} {\bibfnamefont {M.~J.}\
  \bibnamefont {Hartmann}},\ }\bibfield  {title} {\bibinfo {title} {Dissipative
  optomechanical preparation of macroscopic quantum superposition states},\
  }\href {https://doi.org/10.1103/PhysRevLett.116.233604} {\bibfield  {journal}
  {\bibinfo  {journal} {Phys. Rev. Lett.}\ }\textbf {\bibinfo {volume} {116}},\
  \bibinfo {pages} {233604} (\bibinfo {year} {2016})}\BibitemShut {NoStop}%
\bibitem [{\citenamefont {Breuer}\ and\ \citenamefont
  {Petruccione}(2007)}]{Breuer2007}%
  \BibitemOpen
  \bibfield  {author} {\bibinfo {author} {\bibfnamefont {H.-P.}\ \bibnamefont
  {Breuer}}\ and\ \bibinfo {author} {\bibfnamefont {F.}~\bibnamefont
  {Petruccione}},\ }\href@noop {} {\emph {\bibinfo {title} {The Theory of Open
  Quantum Systems}}}\ (\bibinfo  {publisher} {Oxford University Press},\
  \bibinfo {address} {Oxford},\ \bibinfo {year} {2007})\BibitemShut {NoStop}%
\end{thebibliography}%

\end{document}